\documentclass[useAMS,usenatbib]{mn2e}

\usepackage{amsmath}	
\usepackage{graphicx}
\usepackage{lscape}
\usepackage{color}

\title[Box/Peanut/Bar SDSS Nearby Galaxies. I.]{Box/Peanut and
Bar structures in edge-on and face-on SDSS nearby galaxies I. Catalogue}

\author[A. Yoshino and C. Yamauchi]{Akira Yoshino
$^{1}$\thanks{E-mail:yoshino.akira@jaxa.jp (AY)} and Chisato Yamauchi
$^{1}$ $^{2}$ $^{3}$\thanks{E-mail:cyamauch@ir.isas.jaxa.jp (CY)} \\
$^{1}$JAXA (Japan Aerospace Explanation Agency), ISAS (Institute of Space and
Astronautical Science), 3-1-1 Yoshinodai,\\
Sagamihara, Kanagawa, 252-5210, Japan\\
$^{2}$Kimino-cho Misato Astronomical Observatory, 180 Matsugamine,
Kimino-cho, Kaiso-gun, Wakayama, 640-1366, Japan\\
$^{3}$Institute for Education on Space, Wakayama University, 930
Sakaetani, Wakayama, 640-8510, Japan}

\begin{document}

\date{Accepted 2014 October 24. Received 2014 October 20; in original
form 2014 February 28}

\pagerange{\pageref{firstpage}--\pageref{lastpage}} \pubyear{2014}

\maketitle

\label{firstpage}

\begin{abstract}

 We investigate Box/Peanut and bar structures in image data of edge-on
 and face-on nearby galaxies taken from SDSS (Sloan Digital Sky Survey)
 to present catalogues containing the surface brightness parameters
 and the morphology classification. About 1700 edge-on galaxies and 2600
 face-on galaxies are selected from SDSS DR7 in g, r and i-band. The
 images of each galaxy are fitted with the model of 2-dimensional surface
 brightness of S\'{e}rsic bulge and exponential disk. After removing
 some irregular data, Box/Peanut, bar and other structures are
 easily distinguished by eye using residual (observed minus model)
 images. We find 292 Box/Peanut structures in the 1329 edge-on samples
 and 630 bar structures in 1890 face-on samples in i-band, after
 removing some irregular data. Then the fraction of Box/Peanut galaxies
 is about 22 percent against the edge-on samples, and that of bar is
 about 33 percent (about 50 percent if 629 elliptical galaxies are
 removed) against the face-on samples. Furthermore the strengths of the
 Box/Peanut and bar are evaluated as ``Strong'', ``Standard'' and
 ``Weak''. We find that the strength increases slightly with increasing
 $B/T$ (Bulge-to-Total flux ratio), and that the fraction of Box/Peanut
 is generally about a half of that of bar irrespective of the strength
 and the $B/T$. Our result supports the idea that the Box/Peanut is bar
 seen edge-on. 

\end{abstract}


\section{Introduction}

A Box/Peanut (hereafter B/P) structure of disk galaxy, also called boxy
bulge, is known as a galactic morphology with 2-dimensionally axial
symmetric and vertically extended shape like a box or peanut around the
galactic center \citep{b17,b21,b27,b6}. It is seen only in edge-on disk
galaxies. Since the surface brightness of B/P part is generally fainter than
those of bulge and disk, it is not conspicuous except for some famous
galaxies (e.g., NGC 128 \citep{b4}, NGC 2654 and NGC 4469 \citep{b13}). 

Previously several authors have discussed the origin and dynamics of
B/P. The theoretical studies are classified roughly as the following
three ideas. Hypothesis 1: The B/P is a bar; the B/P is the side-on view of the
bar having a vertical shape of ``figure of infinity''
\citep{b9,b5,b21,b22,b2,b27,b8,b37}. 
Hypothesis 2: The B/P is a remnant of infalling satellite galaxy to the
host galaxy. When the satellite falls on the host at an oblique
angle, the tidal force makes the orbits of satellite stars be the
``figure of infinity'' around the host galaxy and then is seen as B/P,
while it yields a polar-ring galaxy when the satellite falls at nearly a
right angle \citep{b36,b23}. 
Hypothesis 3: The B/P is a cross sectional view of
cylindrically extended disk formed from vertical orbit resonance of
disk stars. The vertical resonance of disk stars is generally thought to
be associated with the perturbation by bar, but it occur not only in
barred galaxies but also in normal spiral galaxies \citep{b26}. 

Here is a question whether the B/P feature is major or rare in the universe. 
It is important to estimate the fraction (abundance ratio) of the
galaxies having B/P structure against all edge-on spiral galaxies from
observational study, because the expected fraction is significantly
different between these hypotheses. The first hypothesis leads the
fraction of B/P galaxies against all disk galaxies (hereafter B/P
fraction) to be somewhat lower than that of barred galaxies, because the
B/P structure would not be detected when we look the bar from end-on,
i.e., the end-on view of ``figure of infinity'' would not be such a B/P
shape but be a round or elliptical shape as a normal bulge. Therefore
the B/P fraction is thought to be roughly about a half of barred galaxies;
roughly 20 to 30 percent against all edge-on spiral galaxies in this
idea, because the bar (including strong bars (SB) and weak bars (SAB))
fraction in the present universe is about 40 to 50 percent in optical
bands observation (e.g., 43 percent by \citet{b14}, $44\pm7$ percent
by \citet{b24}, 48-52 percent by \citet{b3}). 
The second hypothesis leads to only a few percent of the B/P fraction,
because such case as the elongated shape of satellite galaxy
orbiting around the host galaxy is seen like the B/P is thought to
be rare, as \citet{b23} have already pointed out. 
In the third hypothesis, the cylindrical disk would always look
like the B/P when it is seen edge-on. In this idea the B/P feature is
not concerned with the view angle (side-on or end-on). Thus the B/P
fraction would be relatively high, about the same as that of bar, i.e.,
about 40 to 50 percent. If spiral galaxies (normal and barred) have
commonly the the vertical resonance orbits of disk stars, the B/P
fraction would be significantly higher than that of bar. 

The B/P fraction has not been well established from observation, because
of the difficulty of extracting the B/P from a large number of
data. Since the B/P is generally not obvious and lies on the
bright region of galaxy, it is necessary to carry out an image
processing to enhance the B/P feature. 
The previous observational studies in 1980s to 2000s have claimed
that various B/P fractions; 1.2 percent \citep{b17} to 45
percent \citep{b12,b21}. 
Such various values would be caused by the difficulty of finding the B/P
structure and the quality of data and the difference of the difinition
of B/P. 
\citet{b17} has found 30 B/P galaxies by eye estimation from the
ESO/SERC J sky servey south (declination -18 degree) photographic plates
and has reported that the B/P fraction is 1.2 percent against all disk
galaxies, however the ``all'' disk galaxies include those of various
inclination. 
\citet{b34} has found 23 B/P galaxies by eye estimation from 117 edge-on
galaxies of RC2 (therefore the B/P fraction is about 20 percent) using
the ESO/SERC J or POSS R-band survey plates using isophote contour
maps. 
\citet{b21} has reported that 330 B/P galaxies by eye
estimation taken from 734 edge-on disk galaxies (therefore the B/P
fraction is 45 percent) of RC3 using mainly isophote contour maps of DSS
(Digitized Sky Survey) and partly CCD images obtained from various
telescopes. However, their sample data contain a large number of late-type
(Sc-Sd) galaxies (111 B/Ps in 249 Sc-Sd galaxies), of which bulges are
generally ambiguous in the DSS images. Many of these late-type B/Ps are
``type 3'' in the paper, i.e., ``bulge is close to box-shaped, not
elliptical'', and thus the B/P fraction may be somewhat overestimated. 

We define the B/P feature in this paper as follows: an apparent surface
brightness feature of a disk galaxy seen like a box or a peanut other
than an elliptical bulge component and a disk component, projected on
the celestial sphere, distributing near the galactic center and on the
vertical direction of the galaxy. 
Hence the B/P feature in this difinition can be detected as a residual
of surface brightness when those of bulge and disk are subtracted from
the galaxy image, if exists.

It is important to research the B/P fraction using recent large survey
digital data of which quality is relatively high and homogeneous, such as
SDSS. Furthermore, it is also important to analyze both edge-on and
face-on galaxies with a same process to extract the B/P or bar
structures when we consider the relationship between B/P and bar. 

We use sample data of nearby galaxies taken from SDSS DR7 \citep{b1} 
to research the B/P fraction for edge-on galaxies and that of bar for face-on
galaxies in this paper. The SDSS image data of nearby galaxies has
high quality of deep, homogeneity and resolution enough to research the
B/P or bar in nearby galaxies. 
We carry out image processing to extract the B/P or bar other than bulge
and disk. First, we perform model fitting for each observed image. 
The model is composed with S\'{e}rsic law bulge and exponential law
disk. Then we subtract the model surface brightness from the image to
make the residual images. The B/P, bar, or other faint structures are
obvious in the residual images and these morphology are classified by
eye. 
We present the catalogues of edge-on and face-on galaxies containing the
results of model fitting (decomposition parameters of bulge and disk)
and the morphology. 
Finally we investigate statistically the relation between B/P and bar and
consider which hypothesis is valid for the result. 

The colors and magnitudes, which are clue to investigate the stellar
population and evolution, are also obtained for the sample data. The
detail will be presented in the next paper II for the future. We focus
on the fraction of B/P and bar in this paper. 

This paper consists of the following section: Section 2 describes the
data selection, Section 3 describes the model fitting and image
processing, Section 4 shows the result, Section 5 is the discussion, and
Section 6 is the conclusion.

\section[Data]{Data}

\subsection{Retrieving Data}

The main aim of this study is to estimate the B/P or bar fraction of
galaxies using a large number of image data observed in optical
wavelength band. It is necessary to use data taken from a large area
survey to obtain the unbiased statistics as much as possible. It is
desirable to use more than several hundred samples to research
statistically. Moreover, to analyze the structures of galaxies, it is
necessary to use the images having enough resolution, limit magnitude
and homogeneous quality. To fulfill these requirements, we use the
images of nearby galaxies (redshift is nearly zero; generally smaller
than 0.1) in three optical bands (g, r and i-band) taken from SDSS DR7
data archive (http://www.sdss.org/dr7/). SDSS DR7 consists of imaging
and spectroscopy data in optical five bands (u, g, r, i and z) covering
8,400 square degrees and containing 930,000 galaxies, of which
photometric and spectroscopic data are automatically evaluated and
tabulated in database. The data quality is usually well; the pixel size
is 0.396 arcsec/pixel, the FWHM (Full Width Half Maximum) of PSF (Point
Spread Function) is about 1.2 to 1.5 arcsec, and the zero point
magnitude is about 24 to 25 $\rm[mag/arcsec^2]$ for g, r and
i-band. The images of u and z-band are not used in this paper because of
the large amount of dust absorption in the galaxy in u-band and the
relatively low signal-to-noise ratio (S/N) in z-band. The images of
nearby galaxies have enough large size; the petrosian radii for many of
our samples are about several ten arcsec. These values are enough to
research the B/Ps or bars for nearby galaxies. 

The SDSS catalogue data and image data are available from the above
web site. The photometry data such as magnitude, size, position
angle, classification (star, galaxy, quasar and so on) are available. 
we can easily retrieve the image data of edge-on galaxies and face-on
galaxies using a suitable search condition. The edge-on samples are
chosen using the following search condition in the galaxy category: 

$r<17$ and $err_r<0.1$ and $petroRad_r>10.0$ and $zConf>0.99$ and
$isoB_r/isoA_r<0.25$.

The values of condition $r<17$ (meaning that the r-band total magnitude
for the object is brighter than 17 magnitude) and $petroRad_r>10$ (the
petrosian radius of r-band is larger than 10 arcsec) are chosen to
obtain bright and large size nearby galaxies. These values are empirically decided, i.e.,
the fainter or smaller galaxies cannot be well analyzed in the following
image processing (see Section 3.7). The condition of $err_r<0.1$ is added
to obtain data having photometric error lower than 0.1 magnitude. In
addition $zConf>0.99$ is the condition to select the data of which
redshift confidence is enough high. The $isoB_r$ and $isoA_r$ are the sizes
of isophotal minor axis and major axis of r-band, respectively. Thus the
$isoB_r/isoA_r$ represents the axial ratio of the galaxy. We define
``edge-on'' as the axial ratio is lower than 0.25 in this paper. We
cannot know the intrinsic thickness of the galaxy and the apparent
inclination beforehand. The 0.25 of axial ratio may be somewhat loose to
select only edge-on galaxies, because the spiral structure is
occasionally appeared for some galaxies, which indicates that the
galaxies are not entirely edge-on but slightly inclined. 
However, if we set the axial ratio to smaller value (for example, 0.2),
some galaxies having large size of B/P are rejected. Note that our
edge-on samples include slightly inclined galaxies. 

On the other hand, the face-on samples are chosen using the search
condition as follows: 

$r<17$ and $err_r<0.1$ and $petroRad_r>10.0$ and $zConf>0.99$ and $isoB_r/isoA_r>0.8$.

The search condition is the same as for edge-on galaxies except for the
axial ratio; larger than 0.8. In addition, only one-tenth data are
extracted randomly, because the number of face-on samples is
significantly larger than that of edge-on samples. 

The numbers of sample data using above condition are 1716 for edge-on
galaxies and 2689 for face-on galaxies.

\section[Data Processing and Analysis]{Data Processing and Analysis}

\subsection{Reduction}

We retrieve ``fpC'' FITS files from SDSS archive using the search
condition as mentioned above section. The fpC data are rectangular
images divided according to numbers of stripes and runs in
SDSS. They are already well reduced, that is, flat fielding has been
done to them. The data reduction that we must carry out before model
fitting is as follows: 1. Subtracting bias level of 1000 ADU from the fpC
images. 2. Cutting roughly these images around each target
galaxy. 3. Rotating each image so that the major axis of galaxy is along
to x-direction using the value of position angle taken from SDSS
database. 4. Shifting the rotated images so that each galactic center is
located at the center of image; $(x, y) = (0, 0)$, where the galactic
center is regarded as the most luminous pixel around the central (bulge)
region. 5. Estimating the S/N value at each pixel for each image. Field
data (PSF FWHM, air mass, zero-point, and so on) for each FITS file are
also obtained from SDSS catalogue. 6. Cutting off the margins of each
image (upper, lower, right and left side) so that the resulting image
has the S/N higher than 3.0. 

\subsection{Model of Surface Brightness}

In general, it is not easy to make a model for the B/P itself because of
the faint flux compared to bulge and disk. Since the main aim in
this paper is the number counting of samples having B/P, it is not
necessary to make the B/P model. 
As \citet{b29} and other authors have already pointed out, various
structures like as spiral arms, bar, nuclear, dust lane and so on other
than bulge and disk are easily revealed in the residual image; the
observed image minus the fitted model image. Therefore, instead
of modeling B/P structure, we find by eye whether B/P (or bar) exists on
residual images other than bulge and disk or not. To obtain the residual
images for each galaxy, we make a surface brightness model of bulge and
disk. We assume that the galaxy consists of bulge and disk, which are
two main components of galaxy. Then we subtract the model from the
observed one to yield the residual image. 

Our model and decomposition method follows \citet{b7}. 
The surface brightness of galaxy is assumed to be composed with S\'{e}rsic
law bulge \citep{b33} and exponential law disk \citep{b16}. We
use the following 2-dimensional model of surface brightness: 

\begin{equation} 
 I_{bulge}(x, y) = I_e*\exp( ( \sqrt{x^2 + (y/(b/a)_B)^2} /r_e) -1 )^{\beta},
\end{equation}
\begin{equation} 
 I_{disk}(x, y) = I_0*\exp(\sqrt{x^2 + (y/(b/a)_D)^2} /r_s ),
\end{equation}
\begin{equation} 
 I_{galaxy}(x, y) = I_{bulge}(x, y) + I_{disk}(x, y),
\end{equation}
\begin{equation} 
 I_{model}(x, y) = \int PSF * (I_{galaxy}(x, y) + const.sky) dxdy,
\end{equation}

where $(x, y)$ is a pixel coordinates from a galactic center, $I_e$ and $r_e$
are an effective intensity (flux) of S\'{e}rsic law in ADU/pixel unit
and an effective radius in pixel unit for bulge, respectively, and $\beta$ is
S\'{e}rsic law index. Moreover $I_0$ and $r_s$ are a central intensity and a
scale length for disk, respectively. $(b/a)_B$ and $(b/a)_D$ are the
axial ratios for bulge and disk, respectively. The model surface brightness of a
galaxy $I_{galaxy}(x, y)$ is simply the sum of those of bulge and disk. We
obtain sky values as the mode of the fpC image and add the values to the
$I_{galaxy}(x, y)$ for each target galaxy. Finally we convolve the value
with Gaussian kernel PSF; the FWHM values for each image are taken from SDSS
archive. 

The seven constant values (bulge $I_e$, bulge $r_e$, bulge $(b/a)_B$, bulge
S\'{e}rsic index $\beta$, disk $I_0$, disk $r_h$ and disk $(b/a)_D$) for each galaxy must
be obtained from the model fitting. When $\beta = 1/4$, the profile is
called as de Vaucouleurs law, which represents the surface brightness
distribution of typical elliptical galaxies, while $\beta = 1$, it is
the same as the exponential law. 

Several authors (e.g., \citet{b30}) have been presented
three-dimensional disk models. 
Moreover a decomposition of disk into a thin disk and a thick disk may
be suitable for some galaxies. 
However, the elaborate model needs generally more parameters to
represent the structure of galaxy than simple model. 
The parameters used in the three-dimensional and thin/thick disk model
are often sensitive to various local structures or asymmetry of the
surface brightness. 
Hence a process of smoothing the image before the model fitting are
needed to reduce the influence in \citet{b30}, however such a process
may introduce an additional error for detecting B/P or bar. 
Furthermore considerably high S/N image are needed to fit with the
elaborate model, however our sample data do not often have such a high
quality. 
We consider that the fitting with the three-dimentional model or the
decomposition of thin/thick disks are valid only for ideal cases of
nearly symmetry edge-on galaxies and having high S/N image. 
On the other hand, the two-dimentional disk model used in this paper is
flexible to represent the disk surface brightness projected on the
celestial sphere, regardless for edge-on or face-on. 
The main theme is to count B/P or bar structures using an uniform
method, hence the two-dimentional model is sufficient for the purpose.

\subsection{Fitting Method}

We use a simple fitting program developed originally to decompose
the 2-dimensional surface brightness into bulge and disk for each
galaxy. Although several excellent softwares of model fitting,
cf. ``galfit'' package \citep{b29} of IRAF, have been
open to the public, we use the original fitting program because of the
simplicity (only seven parameters for two components of bulge and disk)
and the ease of customization for our aim of analyzing thousand data. 

Our fitting program is based on the algorithm of Levenberg-Marquardt
method taken from ``Numerical Recipes in C'' \citep{b31} and
modified it to 2-dimensional version (see also \citet{b38}). 

Prior to the model fitting, the values of flux in the observed
galactic image are read from the FITS file to a text format file in ADU
unit for each pixel: $I_{obs}(x, y)$. The parameter values of model
surface brightness for each galaxy are calculated from a least square
fitting, as minimizing totally $\chi^2$ (chi-square)), which is the sum
of square of difference between the observed values and those of
model. The $\chi^2$ is defined as follows: 

\begin{equation} 
 \chi^2 = \Sigma{( I_{obs}(x, y) - I_{model}(x, y) )^2 / weight^2},
\end{equation}

where the weight is set to the noise level of each pixel;
$\sqrt{I_{obs}(x, y)/gain}$. The ``gain'' value is also taken from SDSS
archive. 

The $\chi^2$ is firstly calculated using initial values of parameters
and then the gradient of $\chi^2$ in the parameter space is
calculated. The resulting parameter values are obtained as the
solution of the matrix of partial differential equations of $\chi^2$
against each parameter. Then the process is repeated,
substituting the resulting parameter values for the previous values. The
fitting is terminated when the $\chi^2$ becomes a minimum value. The
reduced $\chi^2$, which is defined as $\chi^2$ / degree of freedom, is
thought to be an index of success of fitting. In general, it is expected
to be about 1 when the model is well fitted with the observation. 

Dust lane, foreground stars, neighborhood galaxies and so on often
overlap on the target galaxy. 
It is preferable to mask such regions if we would be like to obtain the
pure surface brightness of the galaxy. 
However, the area and the criterion to be masked should be flexible
according to various cases. 
It is not easy to apply automatically a simple and flexible masking
method to various cases in a large number of data. 
Thus the regions of dust lane, foreground stars or neighborhood galaxies
are not masked in this paper.

\subsection{S\'{e}rsic Law Index $\beta$ of Bulge}

The S\'{e}rsic law index $\beta$ of bulge in equation (1)
defines the sharpness of bulge profile on galactic center, and it is
generally thought to be about 1/4 to 1. In general, fine resolution and
small size PSF are needed to define the bulge index $\beta$ from the
model fitting. However, many of apparent sizes of bulges in our sample
galaxies are somewhat small and thus the bulge profiles are affected
considerably by a PSF of seeing. After our first trial, we found that it is
difficult to obtain simultaneously the seven parameters including the
bulge index $\beta$ from the model fitting, i.e., the fitting iteration are not
converged within several ten times iteration of our fitting program and
tend to become unreasonable values. 

However, if the index $\beta$ is fixed to an appropriate value beforehand, the
other six parameters tend to converge to suitable values by only a few
times iteration. Therefore the index $\beta$ is set to any of the following
four constant values; 1.0, 0.75, 0.5 and 0.25. They are chosen as
typical values to represent the bulge surface brightness. Then the other six
parameters are obtained in the fitting. Then we find that the best model
with the minimum $\chi^2$. 

The value of $\beta$ is decided for i-band data, in which the seeing is
usually better and the dust absorption is smaller rather than other band
data, i.e., we firstly analyze the i-band data to decide the index $\beta$,
and then we analyze g- and r-band data applying the $\beta$ value. 

\subsection{Inputting Initial Values}

It is necessary to input initial values to the parameters in equation (1)
and (2) when the fitting program starts. When the initial values are
appropriate (i.e., close to the final solution), the probability of the
success of fitting will be high. Otherwise, the parameters may not
converge or become unreasonable values (extremely large, extremely
small, infinity or lower than zero). Thus we prepare initially forty
template models having suitable values of parameters as described
below. We expect that any of these models will resemble to the target
galaxy and thus it will fit successfully. We apply sequentially these
templates for each sample galaxy in the fitting script to search a best
set of parameter values. 

Assuming $\beta=1.0$, the other initial parameter values are randomly
selected in the following range: \\
Bulge effective intensity $(I_e)_B$ : 5 to 50 ADU/pixel (corresponding
to $(\mu_e)_B$ of 21.7 to 23.8 $\rm mag/arcsec^2$ in i-band), \\
Bulge effective radius $(r_e)_B$ : 2 to 10 arcsec, \\
Bulge axial ratio $(b/a)_B$ : 0.3 to 1.0, \\
Disk central intensity $(I_0)_D$ : 10 to 100 ADU/pixel (corresponding
to $(\mu_0)_D$ of 20.9 to 23.2 $\rm mag/arcsec^2$ in i-band), \\
Disk scale length $(r_s)_D$ : 5 to 30 arcsec, \\
Disk axial ratio $(b/a)_D$ : 0.1 to 0.3, \\
where the $(\mu_e)_B$ and $(\mu_0)_D$ are the SDSS asinh magnitudes
corresponding to the $(I_e)_B$ and $(I_0)_D$, respectively. The
transformation method is described in the Section 3.6. 

The initial models assuming the other S\'{e}rsic index values
($\beta=0.25$, 0.5 and 0.75) and other bands are also made. To match the real
observation, these values are slightly varied as the index $\beta$
(i.e., the $(I_e)_B$ value is set to lower for the lower $\beta$) and bands. 

We also make the initial template models of face-on galaxies. The
initial models for face-on galaxies are equal to those for edge-on
galaxies except for the disk axial ratio (b/a), which is set to be 0.7
to 1.0 instead of the 0.1 to 0.3 for edge-on. 

\begin{figure}

\includegraphics[width=90mm]{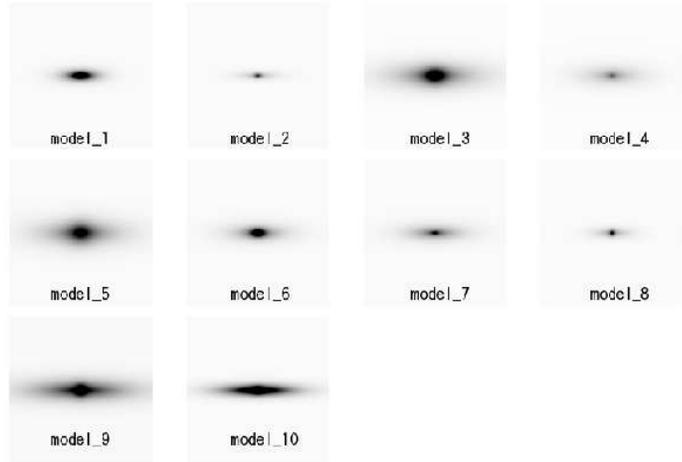}

\caption{Ten template models for the fitting to edge-on galaxies.}
\label{fig1}

\end{figure}

\begin{figure}

\includegraphics[width=90mm]{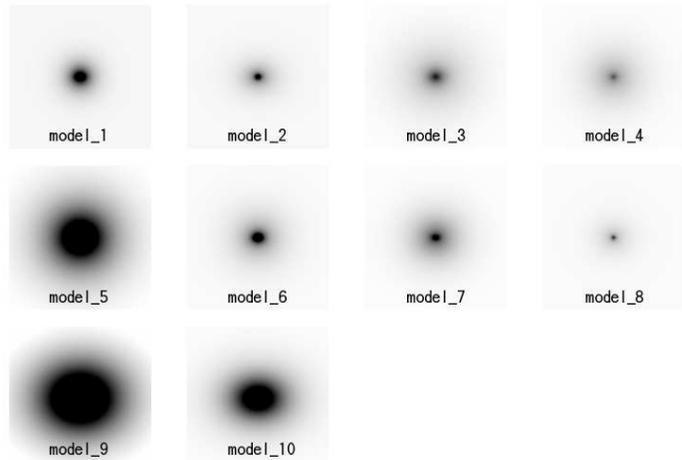}

\caption{Ten template models for the fitting to face-on galaxies.}
\label{fig2}

\end{figure}

Fig.\ref{fig1} shows the ten initial models for edge-on. In
addition Fig.\ref{fig2} shows those for face-on. Finally we prepare
40 initial models (10 models for each bulge index
$\beta=0.25$, 0.5, 0.75 and 1.0) for edge-on, and also 40 models for
face-on. 
In fact, the bulge models of $\beta=1.0$ or 0.75 are well fitted rather
than those of $\beta=0.25$ or 0.5 in many cases of our samples. 

We empirically find that the Levenberg-Marquardt method is considerably
flexible for the input initial values, especially for apparently bright
and large size galaxies. However, the fitting for faint or small size
galaxy is not successfully finished when the initial model galaxy is
considerably different from the target galaxy. We do not carry out more
trials than the above forty initial models for such a faint or small
galaxy because we cannot waste the machine running time. However the
40 models are generally suitable to obtain the fitting results in
many cases. 

\subsection{Transformation from Flux to Magnitude}

The value of effective intensity (flux) $(I_e)_B$ of bulge surface
brightness in ADU/pixel unit is transformed to the corresponding
magnitudes $(mu_e)_B$ in $\rm{mag/arcsec^2}$ unit using the following
algorithms of ``SDSS asinh magnitude system'': 

\begin{equation}
 f/f0 = (((I_e)_B/PS^2)/exptime) * 10^{(0.4*(aa+ kk*airmass))},
\end{equation}
\begin{equation}
 (mu_e)_B = -(2.5/\ln(10)) * [asinh(f/f0)/(2b) + \ln(b)],
\end{equation}

where $f/f0$ is a count rate (count/second), $PS$ is the pixel scale 0.396
(arcsec/pixel), $exptime$ is the exposure time 53.9 (seconds), $aa$ is the
zero point magnitude, $kk$ is the extinction coeffient, $airmass$ is the
degree of airmass, $b$ is a asinh softening coefficients taken from SDSS
archive. The $(mu_0)_D$ of disk central intensity in
$\rm{mag/arcsec^2}$ unit is also calculated from the $(I_0)_B$ instead of
the above $(I_e)_B$. 

The integrated flux $F_B$ for bulge is also calculated from the
model surface brightness of bulge; 

\begin{equation}
 F_B = \int I_B(x, y) dxdy,\\
\end{equation}

where the integration is for all pixels of the image, $I_B(x, y)$ is the
bulge surface brightness model of equation 1. Using disk $I_D(x, y)$
instead of bulge $I_B(x, y)$, we also obtain the disk integrated
flux $F_D$. 

Using the equation of (6) and (7), we can obtain the integrated bulge
magnitude $m_B$ and that of disk $m_D$. We add the suffix to them to
distinguish the observed band. Then, $(m_B)_g$, $(m_B)_r$ and $(m_B)_i$
mean the bulge magnitude for g, r and i-band, respectively, and
$(m_D)_g$, $(m_D)_r$ and $(m_D)_i$ mean those of disk. The galactic
total (whole) flux $F_T$ is obtained from the simple sum of $F_B$ and
$F_D$, and the magnitude $(m_T)_g$, $(m_T)_r$ and $(m_T)_i$ are also
obtained in the same way. 

We also calculate a bulge-to-total flux ratio $(B/T)$, which is a
useful index to evaluate Hubble type of the galaxy, is easily obtained
as the ratio of integrated flux of both component, i.e.,
$F_B/F_T$. $(B/T)_g$, $(B/T)_r$ and $(B/T)_i$ are those of g, r and i-band,
respectively. 

The values of brightness in this paper are not corrected by the
absorption of Our Galaxy, because the absorption is negligible for the
SDSS samples, which are located in the region far from the galactic
plane of Our Galaxy. 

\subsection{Error Estimation Using Model Galaxy Images}

Error of fitting parameter is the difference between a true value and a
value obtained by the fitting. In general, the error is caused by the
background noise, optical limit, pixel resolution and PSF in the
image. Here we make and analyze various artificial galaxy images to
estimate the error. We regard the standard deviation of scattering
between the input values and the output values for the models as the
typical error. The images are made by various parameter values which
are generated randomly within the range of 4 to 200 ADU/pixel for bulge
$I_e$, 0.5 to 5.0 arcsec for bulge $r_e$, 0.3 to 1.0 for bulge $b/a$, 10
to 500 ADU/pixel for disk $I_0$, 3.0 to 30.0 arcsec for disk $r_s$, and
0.1 to 0.3 for disk $b/a$. 

To mimic the observed images, typical noise level and FWHM (1.2-1.5
arcsec) of PSF are added to the images and then the fitting program is
applied to those images. We make one thousand models for each four
S\'{e}rsic index $\beta$ of bulge model and each band.  Then the images are
analyzed using the same procedures as those for real galaxies. 

The resulting values of parameters obtained from our fitting procedure
are compared with the initial values. The scattering between the input
and the output for the one thousand models is thought to be the error
occurred in a probability. 

Fig.\ref{fig21} in Appendix shows the comparison of
input and output parameters for the models of S\'{e}rsic $\beta=1$ bulge
and r-band. Other cases of bulge models and bands are also investigated
and the results are similar to the Fig.\ref{fig21}. 
We empirically find that the scattering for extremely faint galaxies
having $((\mu_e)_B)_r>23\rm[mag/arcsec^2]$ and $(m_D)_r>17.5\rm[mag]$ are
significantly large. We consider that these data are unreliable and thus
they are removed in the Fig.\ref{fig21}. 
(However the data having either $((\mu_e)_B)_r>23\rm[mag/arcsec^2]$ {\it
or} $(m_D)_r>17.5\rm[mag]$  are relatively well reproduced and thus they
are included.)
After removing such extremely faint models, the standard deviations for
bulge $\mu_e$ and disk $\mu_0$ are about 0.13 $\rm[mag/arcsec^2]$ and
0.13 $\rm[mag/arcsec^2]$, respectively, and those of bulge $r_e$ and
disk $r_h$ are about 1.3 arcsec and 0.8 arcsec, respectively. Those for
the integrated flux in magnitude unit $m_B$, $m_D$ and $m_T$ are about
0.31 mag, 0.27 mag and 0.005 mag, respectively. 
There are some outlier data in the figures. They are mainly caused by
the confusing bulge (assumed $\beta=1$) with disk because both profiles
are exponential. When other $\beta$ values (0.25, 0.5 and 0.75) are
assumed, the outlier data are reduced. If the outlier data are excluded,
the standard deviations are rather small than the values described
above. 
In addition we find that the scattering of bulge axial ratio (Fig.\ref{fig21}(b)) 
is somewhat larger than that of disk (Fig.\ref{fig21}(e)). 
The reason is that the small structure of bulge are
influenced by the noise and PSF rather than disk. 

Overall the errors of model fitting are small within the above ranges. 
For caveat, however, the error would be caused by various factors other
than the image quality (resolution, noise and PSF):
contamination of foreground stars, neighborhood galaxies, errors of
pointing (the position angle of the major axis or the location of the
galactic center) or inappropriate initial template model, and so
on. 
The error due to the contamination or the image processing is large
when the galaxy is faint. Hence we use conservatively the condition
$r<17\rm[mag]$ rather than $r<17.5\rm[mag]$ in this paper. 
Note that the total errors would be larger than the above values. In
practice the results (the obtained parameters and the residual images)
are confirmed by eye for each real sample after the fitting as described
in the following Section 3.9.

\subsection{Elliptical Galaxies}

The sample data include elliptical galaxies, because the morphology is
not considered in the selection condition in Section 2. 
Since the fraction of B/P or barred galaxies against all {\it disk}
galaxies is the main subject in this paper, the elliptical galaxies
should be excluded from our samples. 
In fact, our edge-on samples would contain rarely elliptical
galaxies, because such extremely elongated elliptical galaxies as axial
ratio lower than 0.25 are very rare. On the other hand, the face-on
samples contain considerably the ellipticals. We therefore
should remove the elliptical galaxies from the face-on samples. 

Elliptical galaxies are generally thought to have a simple profile of de
Vaucouleurs law.  However, we find that they could be often also fitted
with our model of S\'{e}rsic law bulge plus exponential disk. Thus,
instead of using model fitting to distinguish ellipticals from
spirals, we identify them by eye on the residual images
described in Section 3.10. In practice we can identify elliptical
galaxies as having an apparently no spiral, no patchy cloud in outer
region, but having a concentrated core and/or inner disk-like ring in
the galactic center. 

\subsection{Data Selection}

After the model fitting, we check the resulting parameter values, the
profiles of observed and fitted model, and the residual images by
eye. Most of data are considered to be well fitted to the models,
however some data are not well fitted (i.e., the resulting parameters do
not converge to reasonable values). These failure of fitting are caused
by the following reasons; irregular shape of galaxy (e.g., merger,
irregular galaxy, dusty galaxy, extremely diffuse galaxy and so on),
some extra components superposed on the target galaxy (bright foreground
stars or neighbor galaxies), or the failure of image processing (mainly
error of the centering due to contamination by bright neighbor sources
near the galactic center). The fraction of failure data is about 1/5
against all sample data. 

\begin{figure}
\begin{center}
\includegraphics[width=75mm]{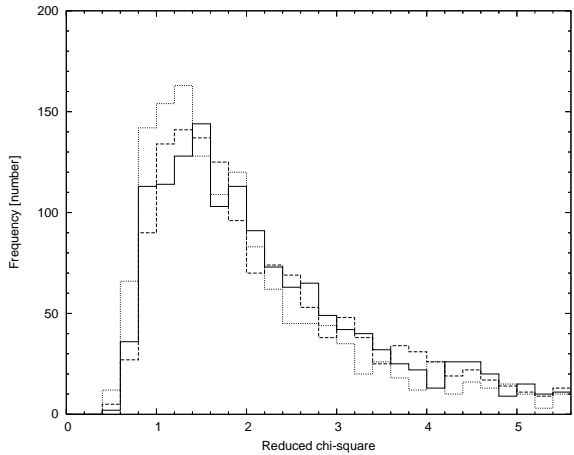}
\caption{Histgram of reduced $\chi^2$ for edge-on data. Solid, dashed
 and dotted lines are those for g, r and i-band, respectively.}
\label{fig3}
\end{center}
\end{figure}

\begin{figure}
\begin{center}
\includegraphics[width=75mm]{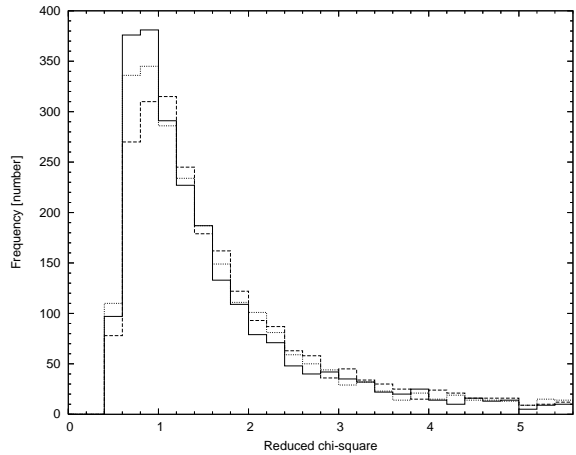}
\caption{Histgram of reduced $\chi^2$ for face-on data. Solid, dashed
 and dotted lines are those for g, r and i-band, respectively.}
\label{fig4}
\end{center}
\end{figure}

Reduced $\chi^2$ (hereafter $\chi^2_\nu$) is one useful index to verify
the fitting result. Fig.\ref{fig3} shows the frequency of the resulting
$\chi^2_\nu$s for edge-on galaxies. Fig.\ref{fig4}  is that of
face-on. These figures indicate that the reduced $\chi^2_\nu$s are
roughly 1 to 3 and that most of our data are well fitted with model. 

However, there are some data having $\chi^2_\nu>>1$. To exclude
these wrong data, it would be appropriate to set an upper limit for the
$\chi^2_\nu$, but the upper limit is ambiguous and arbitrary. In some
cases bright foreground stars let the $\chi^2_\nu$ be large but the
galaxy itself seems to be relatively well fitted to the model. Moreover,
since the B/P or bar is the proper extra components other than bulge and
disk, some right data may be excluded if we set strictly the upper limit
of $\chi^2_\nu$. We therefore empirically set the limit,
checking all sample data by eye on the residual/original images. 

As the result, we find that the limit can be set to relatively
loose. That is to say, the fitting result is regarded as usable in many
cases to find the structures other than bulge and disk even if the
$\chi^2_\nu$ is about 3 to 10. However, if it is larger than 10, the
residual images are not useful to discover the structures and thus the
result is not reliable. 
Therefore we exclude the data of which $\chi^2_\nu$ is higher than
10 in this paper. In addition, we exclude the data having parameter
values of obviously strange; $(b/a)_B$ or $(b/a)_D$ are larger than 5 or
smaller than 0.02; these values of axial ratios are thought to be not
natural. Such unreasonable values of parameters would imply that the
fitting falls into a local minimum of $\chi^2$ and that an another
better model may exist. Due to limited time to analyze, we do not
examine extra other initial models, however. 

We finally select 1253, 1312 and 1329 in g, r and i-band for the edge-on
data and 2042, 2020 and 1890 in g, r and i-band for the face-on data,
respectively. 

\subsection{Residual Images and Visual Classification}

The residual images for each galaxy are produced by the following
process. First, the values of model surface brightness $I_{model}(x, y)$
are produced using the result of fitting. Second, the numeric data of
$I_{model}(x, y)$ in text format file are transformed to the images in
FITS format. Third, we subtract the model image from the observed image
(i.e., subtraction of flux values in ADU unit for each pixel) to produce
the residual image using IRAF 
(version 2.14)\footnote{IRAF is distributed by the National Optical
Astronomy Observatories, which are operated by the Association of
Universities for Research in Astronomy, Inc., under cooperative
agreement with the National Science Foundation.}. 

Finally both the residual and the original images are displayed side by
side using SAOimage DS9 (version 5.2) \citep{b18} with option ``-scale mode
90'' to be inspected by eyes. 

The B/P, bar, spiral arms, rings, satellites, merger remnant and other
structures in the galaxy appear clearly in the residual image
rather than in the original image for each galaxy if they exist, as we
have expected. 
Since we focus on the structures of B/P (edge-on) and bar (face-on) in
this paper, they are classified to sub-class; strong, standard and
weak. We mark them as follows: 

\begin{itemize}
 \item Box/Peanut:
 \begin{itemize}
 \item bx+: Strong Box/Peanut. B/P structure is strong and seen clearly
       like as X-shape on both side of vertical direction. 
 \item bx: Standard Box/Peanut. The boxy shape is obvious but the
       X-shape is not clear.
 \item bx-: Weak Box/Peanut. The boxy shape is not obvious, but a
       rectangular  structure exists along vertical direction other than
       the bulge and disk. The X-shape is not seen. 
 \end{itemize}

 \item Bar:
 \begin{itemize}
 \item br+: Strong bar. Bar structure, overlying on galactic center, is
       equal to or greater than spiral arms. 
 \item br: Standard bar. Bar is obvious but is not outstanding compared
       with spiral arms. 
 \item br-: Weak bar. Bar is not obvious and generally small, but exists
       certainly other than the bulge, disk or spiral. 
 \end{itemize}
\end{itemize}

The other features are also roughly classified. Since we focus on the
B/P and bar, the other features are simplified as long as possible. We
 classify the other features as ``nm'' (normal), ``el'' (elliptical),
 ``ir'' (irregular), ``ov'' (oval), ``rg'' (ring) and ``mg'' (merger),
 and they are not divided into the sub-classes in detail like as
 traditional Hubble-de Vaucouleurs classification (e.g., RSAB(rs)bc). In
 practice, such a elaborated visual classification is appropriate not
 for edge-on galaxies but for face-on galaxies. Instead of the
 elaborated classification, we can infer simply early type or late type
 for each galaxy using the $(B/T)_i$ value; the larger $(B/T)_i$ value
 indicates earlier type. 

\begin{itemize}
 \item nm: Normal disk galaxy (both edge-on and face-on). No B/P or
       bar structure is distinguished. In general, dust lane
       or spiral arms are seen. 
 \item el: Elliptical galaxy (only for face-on). The surface
       brightness is significantly smooth. A bright core and/or an inner
       small disk like as Saturn ring exists generally in the residual
       image. A diffuse ring or shell structure exists occasionally on
       far from the galactic center. Spiral arms or bar are not found. 
 \item ir: Irregular galaxy; including Magellanic type, starburst galaxy
       and significantly disturbed spiral galaxy. 
 \item ov: Oval structure; spheroidal or elliptical component around the
       galactic center, having a smooth surface brightness like as
       elliptical galaxy. The oval resembles a normal bright bulge but
       the structure exists other than bulge in the residual image. It
       also seems to be an intermediate type between bar and bulge. 
 \item rg: Ring structure; including outer ring, inner ring and ``pseudo
       ring'' \citep{b6}. In many cases the ring is combined with oval (or bar) and
       bright companion galaxies. 
 \item mg: Merger. Companion galaxies are conspicuous and are seemingly
       merging into the host galaxy in the image. We do not investigate
       whether a real gravitational interaction between the host and the
       companion exists or not. 
\end{itemize}

When multiple features exist for a galaxy, they are written
side-by-side. For example, ``br-rg'' shows that the galaxy has ``weak
bar'' and ``ring''. Note that the galaxy marked like as ``bx'' or ``br''
has also generally normal bulge, disk and spiral arms. Namely ``nm''
represents that the galaxy has no special features other than bulge,
disk, spiral arms and dust lane. 

We evaluate individually the morphology of galaxy against each
band. In fact, some structure are recognized in g-band but are not
recognized in i-band, and vice versa. The difference of the quality of image
between the bands would give a fluctuation of judgment. Therefore the
classification for a galaxy would vary from g-band to i-band in this
paper. Our classification is not united among the three bands, because
the difference of structures from bluer band to redder band is also
interesting data to research. We discuss the difference in detail in the
following section. 

\begin{figure*}
\begin{center}
\includegraphics[width=135mm]{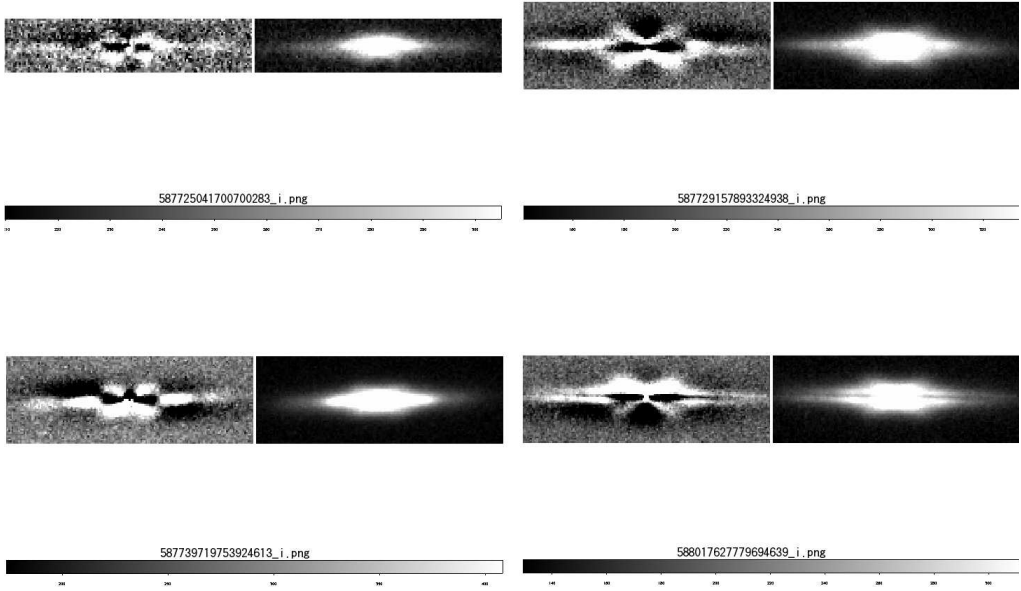}

\caption{Residual (left) and original (right) images for four typical ``bx+''
 (Strong B/P) sample edge-on galaxies in i-band. X-shape structures are obviously
 detected in the residual images. The name in our catalogue for upper-left galaxy:SDSS-eon-0033. Upper-right:SDSS-i-eon-0185. Lower-left:SDSS-i-eon-0753. Lower-right:SDSS-i-eon-1172. }

\label{fig5}
\end{center}
\end{figure*}

\begin{figure*}
\begin{center}
\includegraphics[width=135mm]{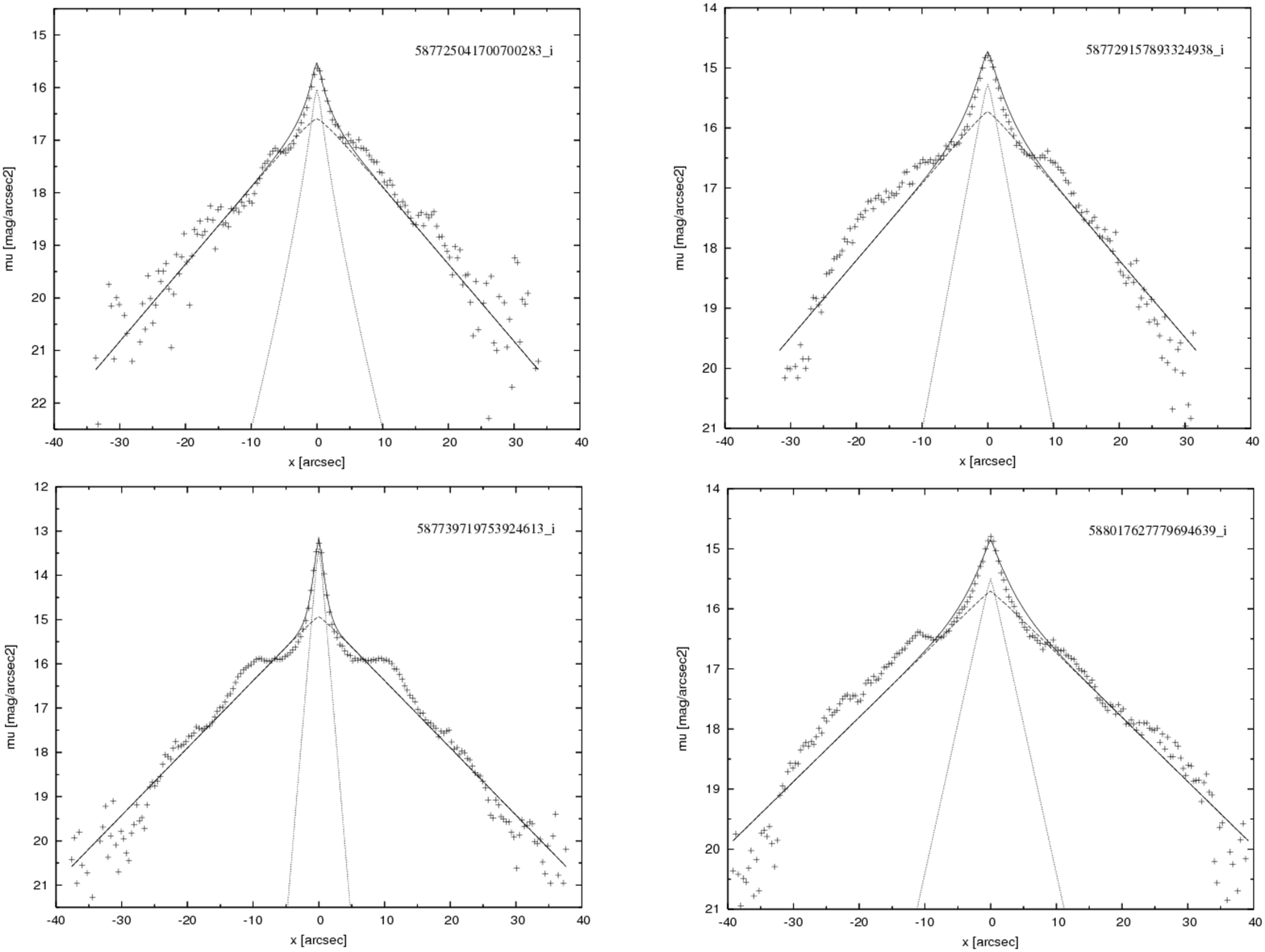}

\caption{The observed and model profiles along the major axes of
 galaxies for the same ``bx+'' edge-on objects as those of Fig.\ref{fig5}. The x-direction is the
 distance from the galactic center along the major axis in arcsec unit,
 while the y-direction is the surface brightness in i-band in $\rm
 mag/arcsec$ unit. Cross points and solid line are the observed profile
 and that of fitted model, respectively. Dotted line and dashed line
 represent the bulge and disk, respectively. }
\label{fig6}
\end{center}
\end{figure*}

\begin{figure*}
\begin{center}
\includegraphics[width=135mm]{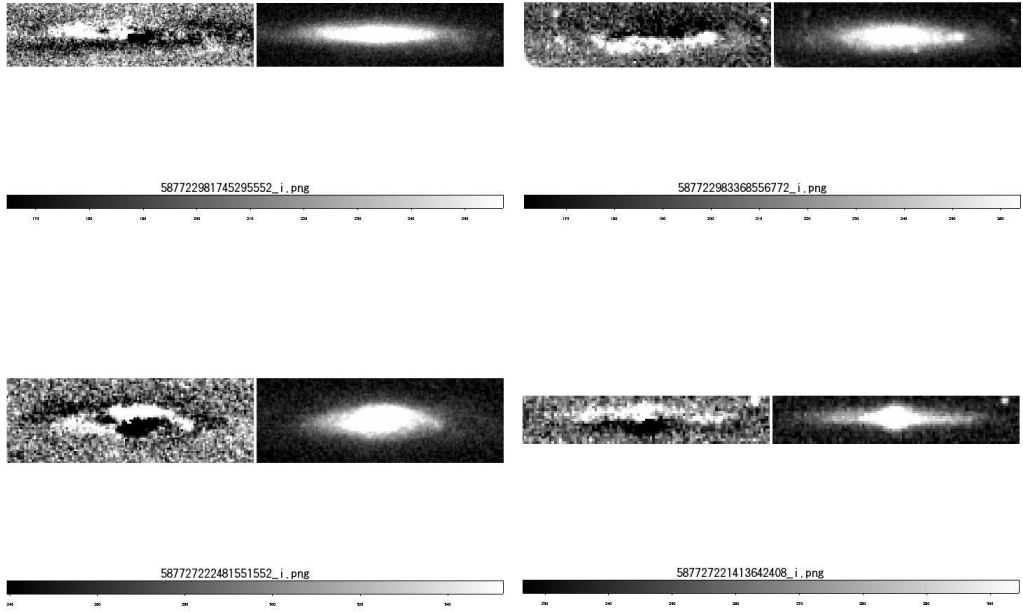}

\caption{Residual (left) and original (right) images for four typical ``nm''
 (Normal) sample edge-on galaxies in i-band. X-shape structure is not found in the
 residual images. The name in our catalogue for upper-left galaxy:SDSS-i-eon-0002. Upper-right:SDSS-i-eon-0009. Lower-left:SDSS-i-eon-0138. Lower-right:SDSS-i-eon-0136. }

\label{fig7}
\end{center}
\end{figure*}

\begin{figure*}
\begin{center}
\includegraphics[width=135mm]{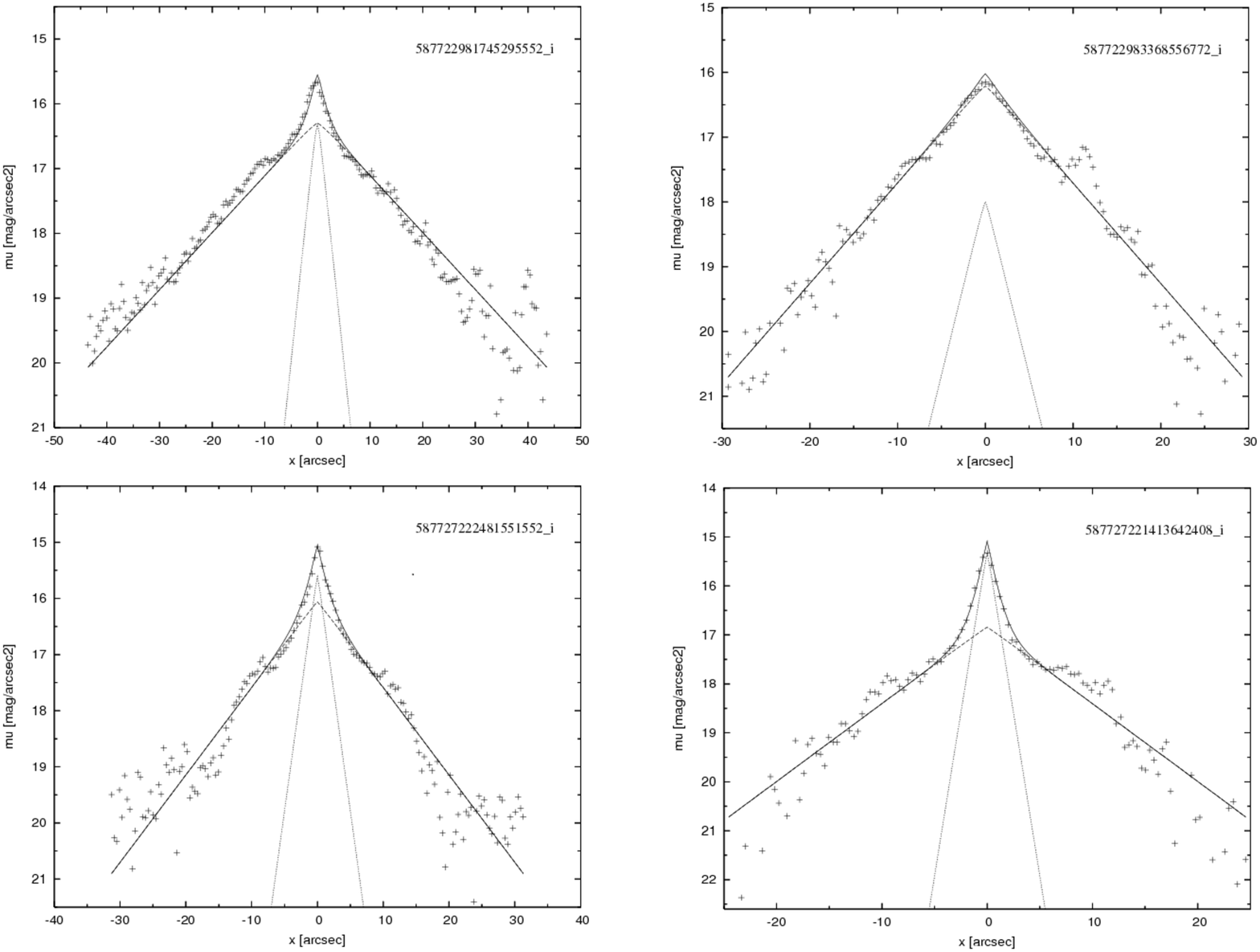}

\caption{The observed and model profiles along the major axes of
 galaxies for the same ``nm'' edge-on objects as those of
 Fig.\ref{fig7}. Both axes, points and lines are the same as those of
 Fig.\ref{fig6}. }
\label{fig8}
\end{center}
\end{figure*}

\begin{figure*}
\begin{center}
\includegraphics[width=135mm]{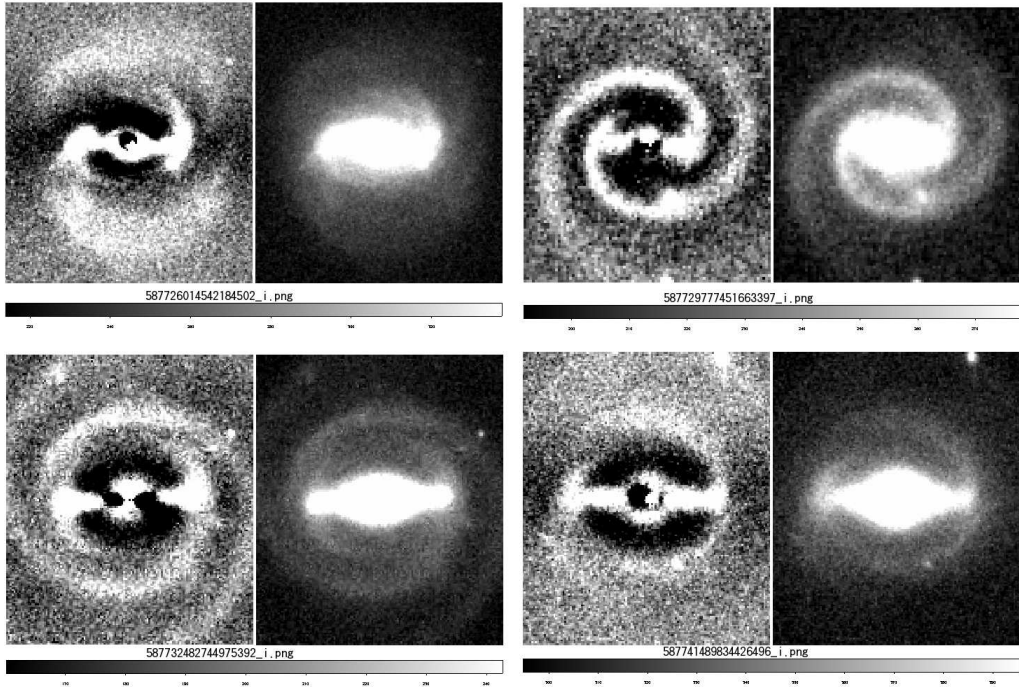}

\caption{Residual (left) and original (right) images for four typical ``br+''
 (Strong bar) sample face-on galaxies in i-band. The name in our catalogue for upper-left galaxy:SDSS-i-fon-0125. Upper-right:SDSS-i-fon-0332. Lower-left:SDSS-i-fon-0494. Lower-right:SDSS-i-fon-1158.}
\label{fig9}
\end{center}
\end{figure*}

\begin{figure*}
\begin{center}
\includegraphics[width=135mm]{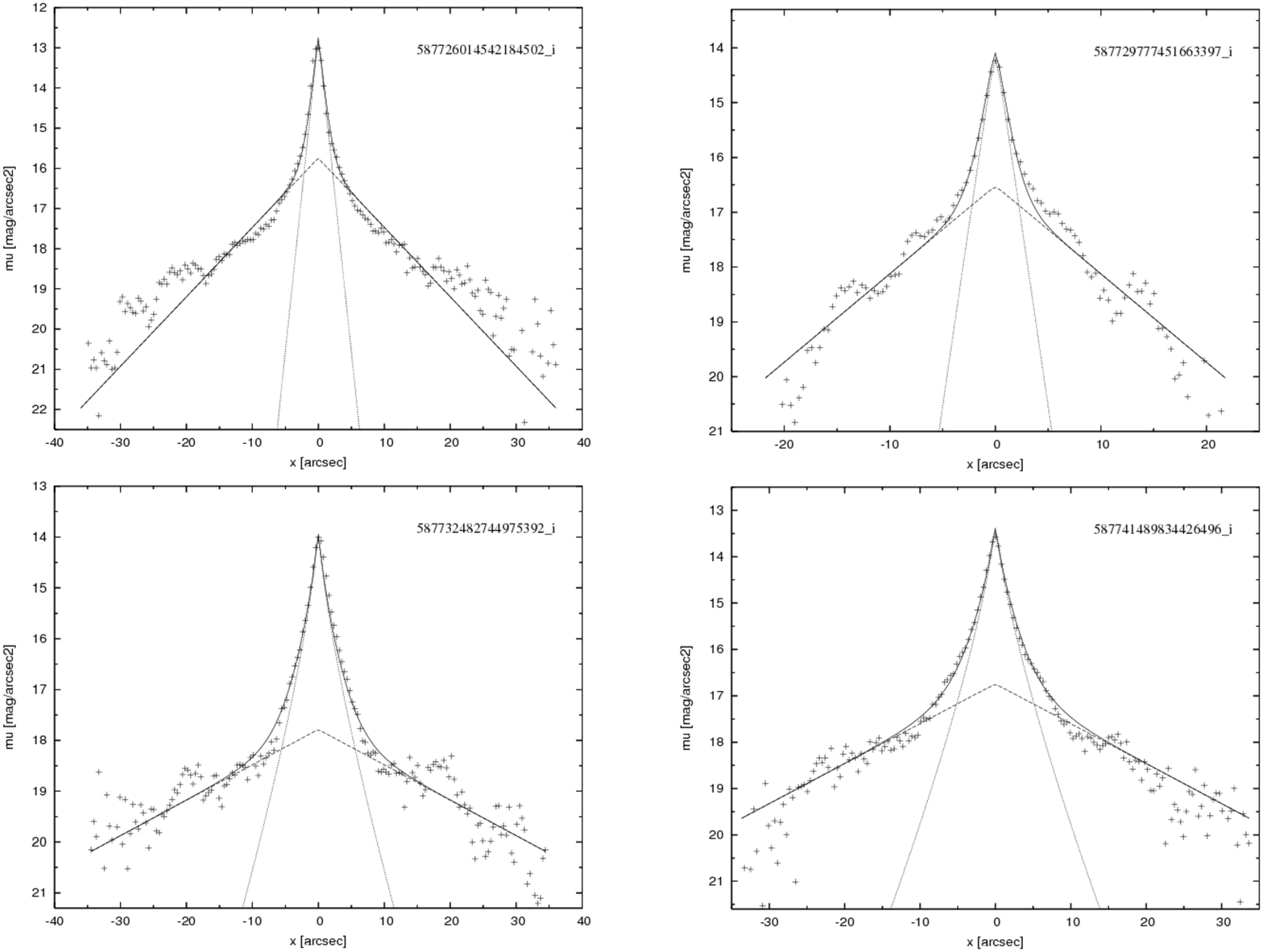}

\caption{The observed and model profiles along the major axes of
 galaxies for the same ``br+'' face-on objects as those of
 Fig.\ref{fig9}. Both axes, points and lines are the same as those of
 Fig.\ref{fig6}. }

\label{fig10}
 
\end{center}
\end{figure*}

\begin{figure*}
\begin{center}
 \includegraphics[width=135mm]{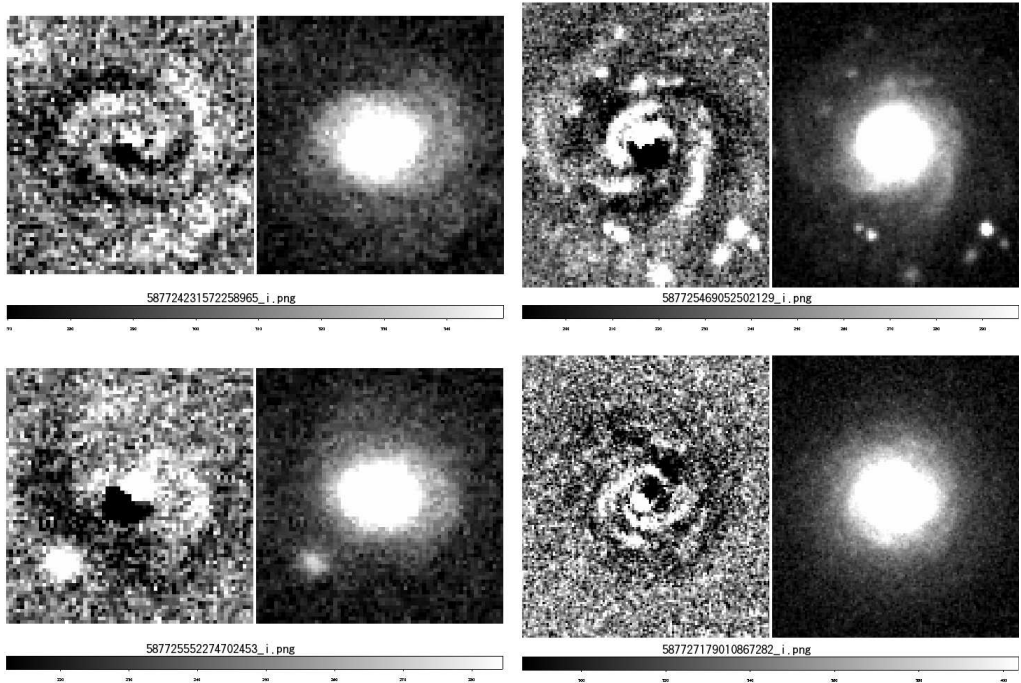}

\caption{Residual (left) and original (right) images for four typical ``nm''
 (Normal) sample face-on galaxies in i-band. The name in our catalogue for upper-left galaxy:SDSS-i-fon-0026. Upper-right:SDSS-i-fon-0063. Lower-left:SDSS-i-fon-0096. Lower-right:SDSS-i-fon-0170.}
\label{fig11}

\end{center}
\end{figure*}

\begin{figure*}
\begin{center}

\includegraphics[width=135mm]{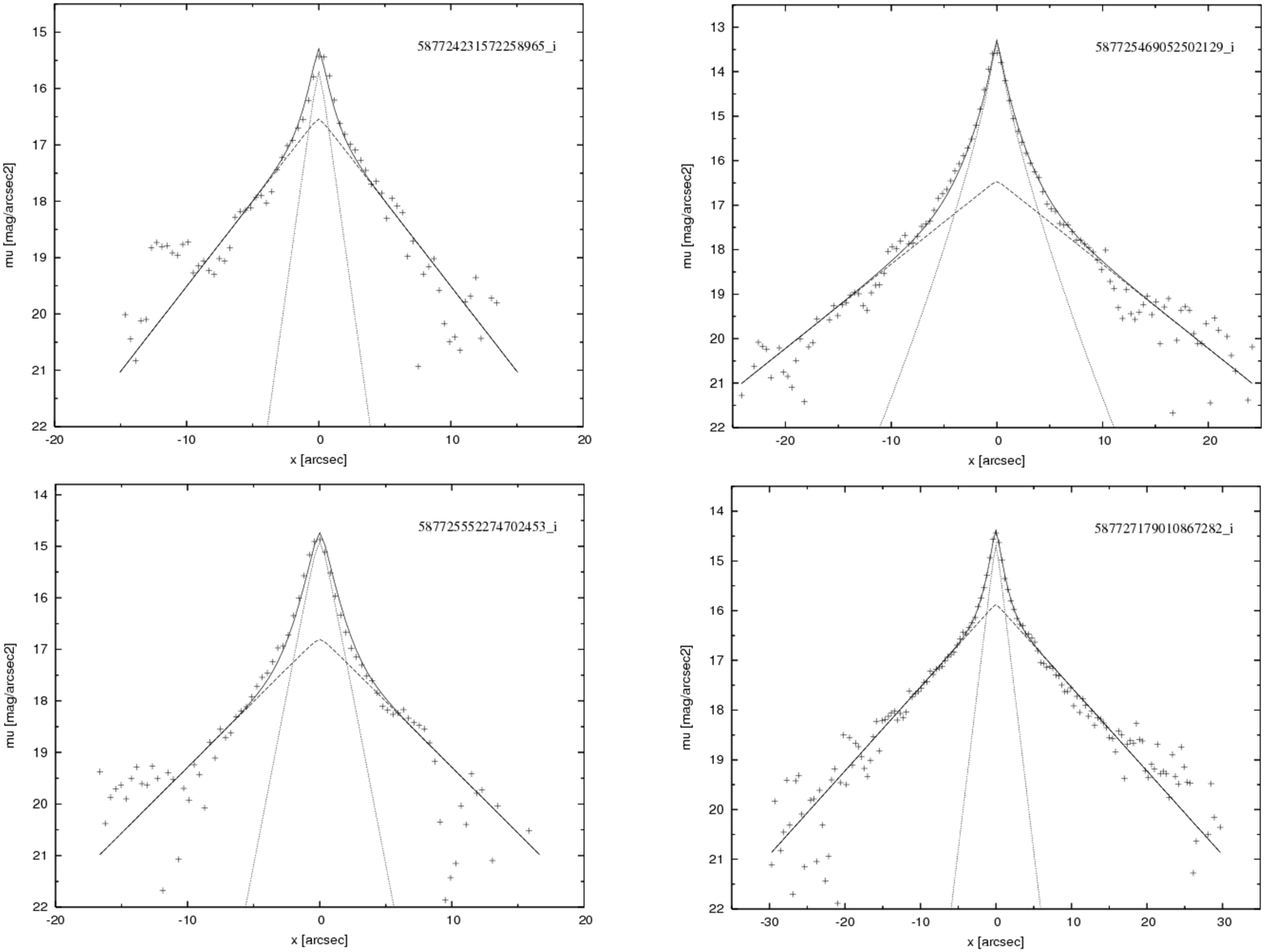}

\caption{The observed and model profiles along the major axes of
 galaxies for the same ``nm'' face-on objects as those of
 Fig.\ref{fig11}. Both axes, points and lines are the same as those of
 Fig.\ref{fig6}. }
\label{fig12}
 
\end{center}
\end{figure*}

Fig.\ref{fig5} shows the residual (left panel) and original
(right panel) images in i-band for four ``bx+'' sample galaxies. 
Fig.\ref{fig6} illustrates the corresponding profiles of observed and
models (bulge, disk and total) along the major axis for each object of
Fig.\ref{fig5}. 
The residual images of each object in Fig.\ref{fig5} clearly show
the B/P features compared to the original images, as we have expected. 
The profiles in Fig.\ref{fig6} indicate that the model fitting are
generally successful. Fig.\ref{fig7} and Fig.\ref{fig8} are those for
four samples of typical ``nm'' (normal) disk galaxies of edge-on. 
Fig.\ref{fig9} and Fig.\ref{fig10} are those for typical ``br+''
(Strong bar) galaxies. 
Fig.\ref{fig11} and Fig.\ref{fig12} are those for typical ``nm''
galaxies of face-on. 

We often find a pair of shoulders on the both sides of profiles for the
edge-on samples in both of Fig.\ref{fig6} and Fig.\ref{fig8}. 
They locate roughly in the end of bulge irrespective of the morphology
(B/P or normal). 
The shoulders of profiles are also found in the face-on galaxies in
Fig.\ref{fig10} and Fig.\ref{fig12}. 
However, they are somewhat weaker than those of edge-on galaxies. Since
the flux density generally increases with the inclination of galaxy, it
is natural that the shoulders of profiles in the edge-on galaxies are
more obvious than those in the face-on galaxies. 
In addition the B/Ps locate generally rather close to the bulge region than
the shoulders in the Fig.\ref{fig6}. 
The lower-left object (SDSS-i-eon-0138) in Fig.\ref{fig7}
is slightly inclined and the spiral arms are clearly distinguished. The
corresponding profile of the lower-left of 
Fig.\ref{fig8} also shows the shoulders. 
Compared to the profiles with the residual images, the shoulders would
coincide with the spiral arms or the end of bar (bar-arm connection)
rather than the B/P or the bar main body. 
Note that the shoulders of profile has been often regarded as the B/P or
bar by previous researchers. Our result shows that the residual images
are useful tool to identify the B/Ps or bars from spiral arms compared
to the original images or the one dimensional profiles. 

For caveat, there exist the error of the center and that of the rotation
of the galaxy image caused by our reduction processing. 
The center is redefined as the brightest pixel around the bulge region,
instead of the center taken from SDSS database. 
On the other hand, the value of position angle used in the processing is
directly quoted from SDSS, i.e., it is not measured with our own way. 
The brightest point is often slightly shifted from the center by SDSS
(within a few pixels) because of a dust lane or the other bump of
brightness. 
Then the major axis of the galaxy is parallel to x-axis (y-direction
shift) in the upper-right of Fig.\ref{fig5}. 
In addition, the shift of center or the other errors in the reduction
would cause an error of rotation angle (within a few degrees). 
Lower-left of Fig.\ref{fig5} 
is the conspicuous case; the major axis of
the galaxy is not consistent with x-direction. 
It is desirable to correct the rotation angle according the amount of
shift of center, but the correction is very slight in most cases. 
Note that we have not carried out the correction of position angle to
simplify the processing. 
However, the model fitting is generally successful and the B/P, bar or
other structures are well detected within the error of shift of a few
pixels and that of rotation of a few degrees. 
Thus we conclude that the influence by the errors on our result is
slight and negligible.

\section{Results}

\subsection{Catalogues of Surface Brightness Parameters and Morphology}

We tabulate the results of fitting and morphology classification
in Table \ref{tab1} and Table \ref{tab2} for edge-on and face-on
samples, respectively. 
Column (1) in the Table \ref{tab1} is an identifier. An example of the
identifier is ``SDSS-i-eon-0001'', meaning ``SDSS data, i-band,
edge-on, and the serial number 0001''. 
That of table \ref{tab2} is ``SDSS-i-fon-0001'', meaning ``SDSS data,
i-band, face-on, and the serial number 0001''.

The objects are ordered by the ``objID'' of SDSS DR7. 
The objID and redshift are quoted from SDSS catalogue and they are shown
in column (2) and (3). 
Columns (4) to (15) are the parameter values of surface brightness
fitting. 
The fluxes are represented in asinh magnitudes unit. 
The effective radius of bulge and the scale length of disk are
represented in arcsec unit. 
The morphology (bx, br, nm, el and so on.) by our eye estimation is
given in column (16). 

The larger $B/T$ indicates the earlier type. 
Here we define the galaxies of $B/T\leq0.05$ as ``Sd'' type, those of
$0.05<B/T\leq0.1$ as ``Sc'', those of $0.1<B/T\leq0.2$ as ``Sbc'', those of
$0.2<B/T\leq0.3$ as ``Sb'', $0.3<B/T\leq0.4$ as ``Sab'', $0.4<B/T$ as ``Sa/E''
(Sa or earlier type galaxies). 
The classification is valid only for this paper but are useful to
consider roughly the relation between the fraction and traditional
Hubble type. 
The Hubble type transformed by the $B/T$ is given in column (17) of the
tables. 

We should note that there are some biases in this study. 
The Hubble type in this paper may be later than that of other studies
because of the following reason: 
The central peak of bulge is influenced by the size of galaxy and the
resolution.
The central bulge profile tends to be mild and to be seemingly high
Sersic $\beta$ (i.e., almost pure exponential) in case of low resolution
(e.g., \citet{b20}). 
The seemingly exponential bulge leads the bulge flux $B$ to small (thus
$B/T$ is small) and then the Hubble type tends to be later. 
In addition, there is a selection effect for edge-on samples, i.e.,
Sa-S0s (generally having large bulge) having axial ratio larger than
0.25 are not included in our edge-on samples. 
Moreover, the elliptical galaxies (32-33 percent of face-on samples)
selected by eye, which are excluded in the following figures, may be
partly the most early type disk galaxies. 
These facts on observation and/or analysis cause the low fraction
of Sa or earlier galaxies in this paper compared to other studies.

We give the another galaxy name taken from ``PGC'' galaxies catalogue
\citep{b28} for each our SDSS object. 
The matching is performed using ``VizieR Service''
(http://vizier.u-strasbg.fr/viz-bin/VizieR) with the search condition
that the corresponding object exists within 5 arcsec radius. 
Column (18) and (19) are the PGC number and the distances from the
position of SDSS object to that of PGC in arcsec unit. 
The distance is generally smaller than 3 arcsec.
The nearest PGC object for each our SDSS object is selected when more
than one objects are found in the PGC catalogue within 5 arcsec radius. 
About 84-85 percent for edge-on data and about 87 percent for face-on
data are matched with the PGC object. 

We also present Table \ref{tab3} and Table \ref{tab4} containing only
B/P or barred galaxies.
They are extracted from Table \ref{tab1} and \ref{tab2}, respectively. 
Table \ref{tab3} is ``The catalogue of edge-on Box/Peanut galaxies
selected from SDSS'', and Table \ref{tab4} is ``The catalogue of face-on
Barred galaxies selected from SDSS''. 

The catalogues of edge-on galaxies includes 1253, 1312 and 1329 objects
for g, r and i-band, respectively, and those of face-on galaxies
includes 2042, 2020 and 1890 objects for g, r and i-band, respectively. 
In addition, the catalogues of B/P galaxies includes 228, 283 and 292
objects for g, r and i-band, respectively, and those of barred galaxies
includes 732, 644 and 630 barred galaxies for g, r and i-band,
respectively. 

Since these tables are significantly large, only the results of 
10 samples for i-band are shown for each table in this paper. The whole
tables will be archived on electronic text format in CDS
(http://cds.u-strasbg.fr/). 
The electronic version include the following additional columns: the
serial number for B/P or bar, the object number using recommended SDSS
format like as ``SDSS JHHMMSS.SS+DDMMSS.S'', the coordinate of RA (right
ascension) and DEC (declination) in epoch J2000.0 taken from SDSS DR7.

\subsection{Fraction of B/P and bar}

The fractions of the galaxies having B/P structure (including bx+, bx and
bx-) against all edge-on sample galaxies are 228/1253 (18 percent), 283/1312
(21 percent) and 292/1329 (22 percent) in g, r and i-band,
respectively. That is to say, the fraction is slightly large in i-band
and small in g-band. The difference would be due to that some B/Ps
embedded deeply in main component (bulge and disk) are easily found in
i-band, of which light absorption by dust is reduced rather than bluer
bands. 

On the other hand, the fractions of the galaxies having bar (including
br+, br and br-) against all face-on sample galaxies are 732/2042 (36
percent), 644/2020 (32 percent) and 630/1890 (33 percent) in g, r and
i-band, respectively. However, the above data contain elliptical
galaxies. The numbers of elliptical galaxies are 671, 640 and 629 in g,
r and i-band, respectively. After removing them, the bar fractions are
732/1371 (53 percent), 644/1380 (47 percent) and 630/1261 (50 percent)
in g, r and i-band, respectively. 

The dispersion of bar fractions between the three bands would be mainly
caused by the difference of S/N and the contamination of ``oval''
feature. It is relatively easy to distinguish the oval from other features
for r-band image, of which the S/N is higher than other bands. If the
S/N is low, however, the oval is sometimes not well identified by eye
and tends to be classified as bar or bulge. The fractions of ovals are not
negligible; 104/1371 (8 percent), 174/1380 (13 percent) and 115/1261 (9
percent) in g, r and i-band data, respectively. Consequently the
bar fractions are not uniform in the three bands and that of r-band is
somewhat lower than those of other bands.

\subsection{Comparison with model magnitudes of SDSS catalogue}

\begin{figure}
\begin{center}
 \includegraphics[width=75mm]{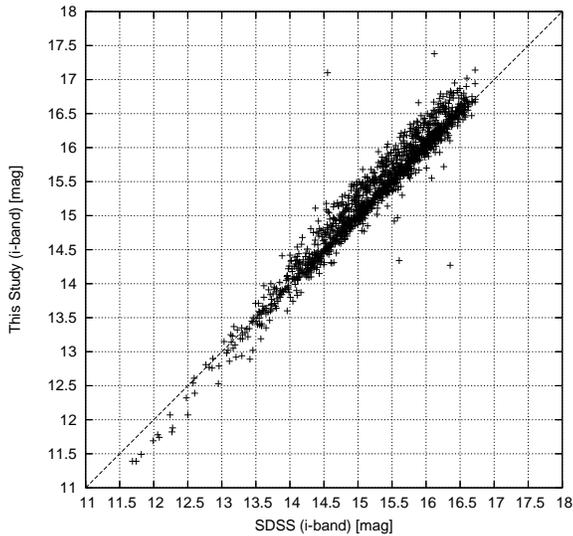}
\caption{Comparison SDSS magnitudes with those obtained from this study
 for edge-on galaxies in i-band.}
\label{fig13}

\end{center}
\end{figure}

\begin{figure}
\begin{center}
 \includegraphics[width=75mm]{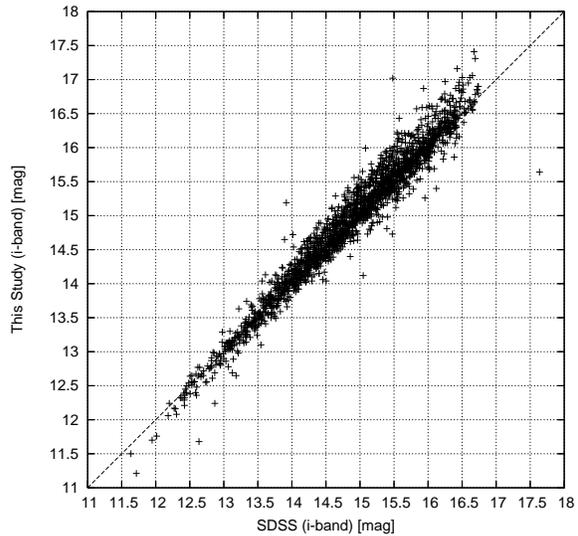}
\caption{Comparison SDSS magnitudes with those obtained from this study
 for face-on galaxies in i-band.}
\label{fig14}

\end{center}
\end{figure}

We compare the resulting model magnitudes (i.e., $(m_T)_g$, $(m_T)_r$ and
$(m_T)_i$) of this study with those of SDSS catalogue to verify the result
of fitting. The method of model fitting adopted by SDSS project is similar to
that of this paper. 
Fig.\ref{fig13} shows the comparison of total
magnitudes of ``modelMag'' taken from SDSS catalogue with the resulting
model magnitudes obtained from this study for the edge-on
samples. 
Fig.\ref{fig14} is that for the face-on samples. 

We find that our results are generally consistent with those of SDSS,
though some scattering and shift are present. First, there is a trend that the
scattering increases with decreasing the brightness (larger magnitude)
in the figure. The scattering would be mainly caused by the background
noise. Second, we find a slight systematic difference; the magnitudes of this study
are slightly brighter than the SDSS values around the bright end
(lower-left) of both figures, while they are somewhat darker around the
faint end (upper-right). The systematic error is probably caused by the
difference of galaxy model and the truncated radius. The model is
composed of S\'{e}rsic bulge and exponential disk in this paper, while
it is a pure de Vaucouleurs law for early type galaxies or a pure
exponential law for late type galaxies in SDSS. Therefore, bulge
luminosities for most of spiral galaxies are not included in the
modelMag of SDSS, but are correctly included in our model. Moreover the
aperture used in the photometry is different; all pixels of the
rectangular area truncated at about $S/N=3$ along the major and minor
axes are used to integrate the flux in this paper, while that of SDSS is
a circular aperture and is truncated at 3 times of the effective radius
for pure exponential profile (or at 7 times of it for pure de Vaucouleurs profile). 
Moreover the model fitting weighted by its flux in our analysis would
also cause the difference. The weighting is suitable to fit to major
bright components, but faint extended disks would tend to be
underestimated. Thus the total magnitudes for faint galaxies would be
somewhat darker than those of SDSS. 

The standard deviations of model magnitudes between SDSS and our result
are 0.22 mag, 0.20 mag and 0.23 mag for the edge-on samples in g, r and
i-band, respectively. Those for the face-on samples are 0.23 mag, 0.20
mag and 0.22 mag in g, r and i-band, respectively. Overall, the results
of model fitting of this study are thought to be generally consistent
with those of SDSS. 

\section{Discussion}

\subsection{Comparison between the B/P Fraction and the Bar Fraction}

As described in Section 4.2, the B/P fractions obtained from our
analysis are about 18, 21, and 22 percent for each g, r and i-band
samples, respectively. On the other hand, those of bar (after
removing elliptical galaxies) are about 54, 47 and 50 percent for each
g, r and i-band samples, respectively. Then the bar-to-B/P ratio is
about 3.0, 2.2 and 2.3 for g, r and i-band, respectively. That is to
say, the B/P fraction is roughly smaller than 1/2 of that of bar. To
consider the relationship between B/P and bar, we should explore the
reason. 

One explanation is that the bar of edge-on view do not always look
like to be B/P, as mentioned in Section 1; the end-on view of ``figure of
infinity'' of bar would seem to be a normal round or elliptical shape as a
bulge. Thus the B/P fraction in edge-on data becomes somewhat lower
than that of bar in face-on data. 
Moreover, note that our edge-on samples include the slightly inclined
galaxies. The B/Ps could not appear in these inclined data, even if they
exist. Thus the contamination of inclined galaxies lets the B/P fraction
be lower compared with the proper value. Therefore it would become
somewhat less than the expected value. Consequently we conclude that at
least our result is consistent with the hypothesis 1 in Section 1;
the B/P is side-on view of bar. 

\subsection{Trend between fraction, strength and band}

We investigate correlations between the fraction, strength and the observed band
for both B/P and bar. Firstly we research the trend between the
frequency of B/P strength against all edge-on samples and the observed bands. The
results are following: 

Fraction of Weak B/P (F(bx-)): 166/1253 (13.2 percent), 189/1312 (14.4
percent) and 155/1329 (11.7 percent) for g, r and i-band, respectively. 

Fraction of Standard B/P (F(bx)): 54/1253 (4.3 percent), 82/1312 (6.3
percent) and 118/1329 (8.9 percent) for g, r and i-band, respectively. 

Fraction of Strong B/P (F(bx+)): 8/1253 (0.6 percent), 12/1312 (0.9
percent) and 19/1329 (1.4 percent) for g, r and i-band, respectively. 

\begin{figure}
\begin{center}
 \includegraphics[width=70mm]{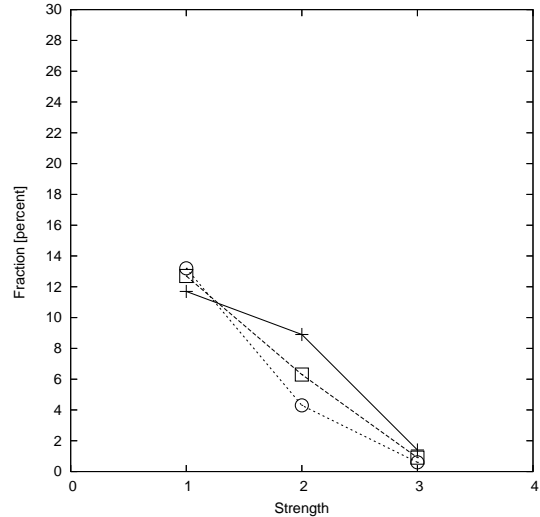}
\caption{Comparison with the B/P strength and the fraction. The x-direction indicates the strength (1=weak, 2=standard, 3=strong). The y-direction indicates the fraction (percentage) against all edge-on samples of each band. The circles, squares and crosses show g, r and i-band data, respectively.}

\label{fig15}

\end{center}
\end{figure}

The result is illustrated in Fig.\ref{fig15}. Two trends are found in
the figure: one is that the fraction decreases with increasing the
strength, the other is that the fraction tends to increase slightly with
the redder band. The reason of the trend would be mainly an absorption
by dust. That is to say, the B/P structure is detected obviously in the
longer wavelength band, in which the dust absorption is small. The B/P
fraction in i-band is thought to be close to the true value. 

The other reason may be the star population, i.e., the stars of B/P may be
constructed with mainly old redder stars and thus the B/P structure is
obvious in the redder band. 

On the other hand, the fractions of the bar against all face-on samples
(excluding ellipticals) are following: 

Fraction of Weak bar (F(br-)): 352/1371 (25.7 percent), 296/1380
(21.4 percent) and 293/1261 (23.2 percent) for g, r and i-band,
respectively. 

Fraction of Standard bar (F(br)): 341/1371 (24.9 percent), 311/1380
(22.5 percent) and 292/1261 (23.2 percent) for g, r and i-band,
respectively. 

Fraction of Strong bar (F(br+)): 39/1371 (2.8 percent), 37/1380 (2.7
percent) and 45/1261 (3.6 percent) for g, r and i-band, respectively. 

\begin{figure}
\begin{center}
 \includegraphics[width=70mm]{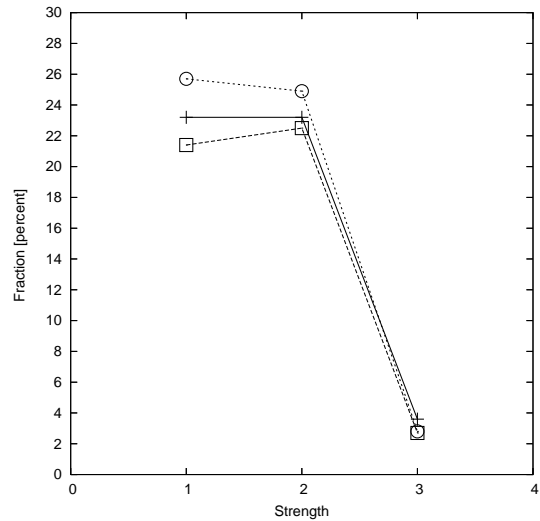}
\caption{Comparison with the bar strength and the fraction. The
 x-direction indicates the strength(1=weak, 2=standard, 3=strong). The
 y-direction indicates the fraction (percentage) against all face-on
 samples (excluding elliptical galaxies) of each
 band. The circles, squares and crosses show g, r and i-band data, respectively.}

\label{fig16}

\end{center}
\end{figure}

The result is illustrated in Fig.\ref{fig16}. 
We find that the fraction decreases with increasing the strength,
however only between F(br) and F(br+). 
Moreover the difference of the bar fraction among the bands is almost
negligible. The reason would be that bars of face-on galaxies would be
seldom hidden by the dust. 

Note that the ``Weak'' include ``apparently unclear'' in addition to
``intrinsically weak''. In fact, the average and the standard deviation of
total (apparent) magnitudes in i-band are 15.22$\pm$0.75 for ``Weak
B/Ps'', that is 14.70$\pm$0.91 for ``Standard B/Ps'' and that is
14.33$\pm$0.63 for ``Strong B/Ps''.  
Moreover, those of bars are 15.25$\pm$0.80, 14.96$\pm$0.89 and
13.96$\pm$0.72 for ``Weak bar'', ``Standard bar'' and ``Strong bar'',
respectively. That is to say, the strength decreases with increasing the
magnitude of galaxy. 
These data show that the weaker structures are partially due to the
lack of S/N of images. 
Thus the F(bx+), F(bx), F(br+), and F(br) obtained from our analysis
should be somewhat underestimated rather than those intrinsic values,
while the F(bx-) and F(br-) should be somewhat overestimated. 

To reduce the effect of dust absorption and S/N, We compare
between the fractions of B/Ps and those of bars in i-band. The results
are following:  F(br+)/F(bx+) = 2.4, F(br)/F(bx) = 2.4, and
F(br-)/F(bx-) = 1.9. 
The fact indicates that the bar-to-B/P abundance ratio for i-band data
is almost constant irrespective to the strength; roughly 2 to 2.5. 
The fact is consistent with the idea that B/P is bar. That is to say, a
strong B/P is thought to be a side-on view of strong bar, a standard B/P
to be that of standard bar, and a weak B/P to be that of weak bar. 

\subsection{The B/P Fraction and the Bar Fraction against $B/T$ (Hubble Type)}

\begin{figure}
\begin{center}
\includegraphics[width=80mm]{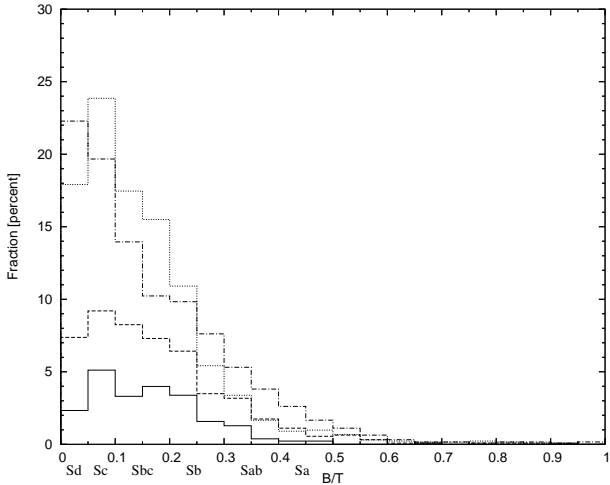}
\caption{Comparison with the $B/T$ and the fractions (percentage) of sample data. The
 x-direction indicates the $B/T$ value. The y-direction indicates the fraction
 (number) of B/Ps (solid line), bars (dashed line), all edge-on (dotted line), and all
 face-on (dot-dashed line) data (excluding elliptical galaxies) of i-band. }

\label{fig17}

\end{center}
\end{figure}

We investigate the F(bx) or the F(br) versus the $B/T$ values 
(Hubble Type). 
Fig.\ref{fig17} shows the comparison between the $B/T$ values
and the fractions of B/Ps, bars, all edge-on and all face-on (excluding
ellipticals) galaxies in i-band. 

We find the following trends in the figure: 

1. The fractions decrease generally with increasing the $B/T$ for each
   B/P, bar, all edge-on, and all face-on. 

2. The peaks lie on the type ``Sc'' for each B/P, bar,
   all edge-on. That for all face-on galaxies lies on the type ``Sd'',
   however the difference of fractions between the two bins is small.

3. The B/P fractions are roughly about a half of the bar fractions
   through the $B/T$.  

\begin{figure}
\begin{center}
\includegraphics[width=80mm]{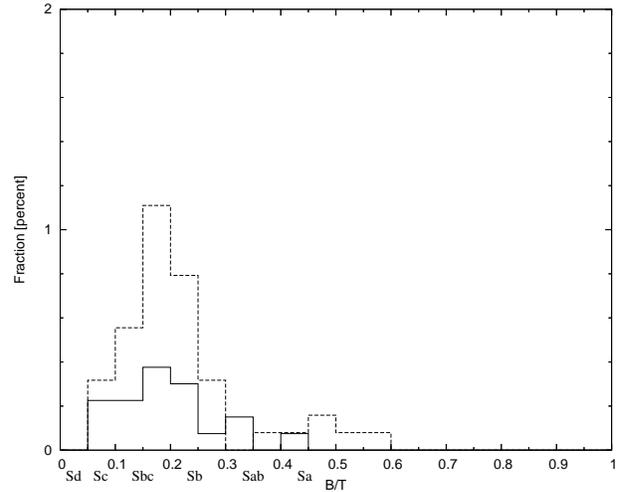}
\caption{The same figure as Fig.\ref{fig17} but only for the ``Strong
 B/Ps'' (solid line) and the ``Strong bars'' (dashed line). }

\label{fig18}

\end{center}
\end{figure}

\begin{figure}
\begin{center}
\includegraphics[width=80mm]{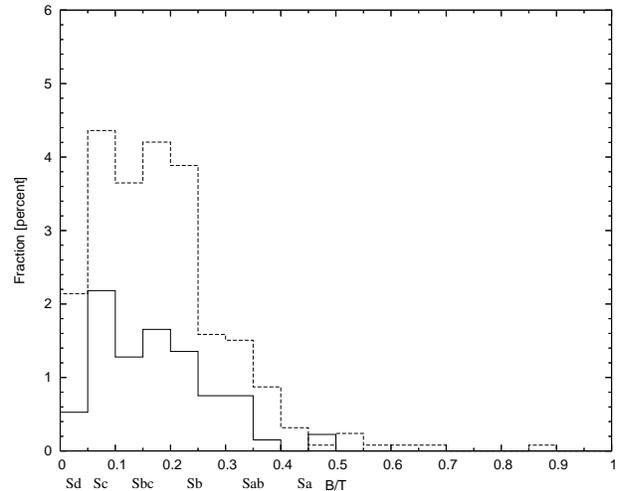}
\caption{The same figure as Fig.\ref{fig17} but only for the ``Standard
 B/Ps'' (solid line) and the ``Standard bars'' (dashed line). }

\label{fig19}

\end{center}
\end{figure}

\begin{figure}
\begin{center}
\includegraphics[width=80mm]{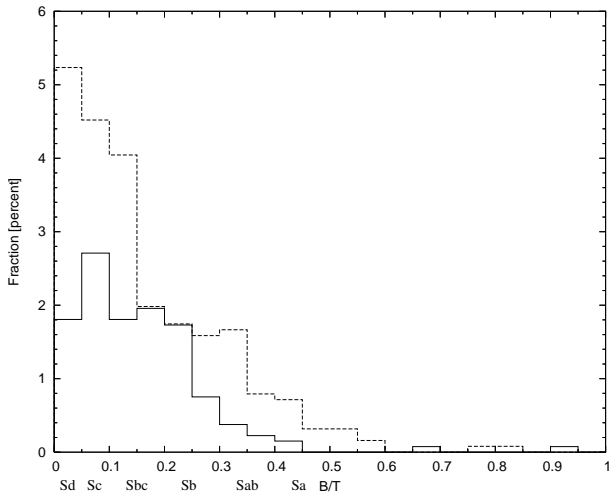}
\caption{The same figure as Fig.\ref{fig17} but only for the ``Weak
 B/Ps'' (solid line) and the ``Weak bars'' (dashed line). }

\label{fig20}

\end{center}
\end{figure}

We also investigate the relation, dividing the fractions into
``Strong'', ``Standard'' and ``Weak''. 
Fig.\ref{fig18} shows F(bx+) and F(br+) versus the $B/T$. 
Fig.\ref{fig19} shows that of F(bx) and
F(br), and Fig.\ref{fig20} shows that of F(bx-) and that of F(br-). 

We find the following trends in the Fig.\ref{fig18} to Fig.\ref{fig20}: 

4. Almost the same trends (peak and gradient) exist both for F(bx+) and
   F(br+), both for F(bx) and F(br), and both for F(bx-) and F(br-). The
   averages and standard deviations of $B/T$ for F(bx+) and F(br+) are
   $0.19\pm0.09$ and $0.22\pm0.12$, respectively. Those for F(bx) and
   F(br) are $0.17\pm0.10$ and $0.18\pm0.12$, and those for F(bx-) and
   F(br-) are $0.15\pm0.12$ and $0.16\pm0.14$, respectively. 

5. The B/P fractions are usually roughly half of those of bars, irrespective
   of the strength or the $B/T$. 

We therefore conclude that the B/Ps have the same trend of fraction as
that of bars over the Hubble type, irrespective of the strength, and
that the abundance of B/Ps in the sample galaxies is roughly a half of
bars. We also conclude that the strength increases slightly with
increasing $B/T$ for both of B/P and bar. We find that both of ``Strong B/Ps'' and
``Strong bars'' lie mainly in ``Sb-Sbc'', both of ``Standard B/Ps'' and
``Standard bars'' lie mainly in ``Sbc-Sc'' (in addition both have double
peaks in the Fig.\ref{fig19}), and both of ``Weak B/Ps'' and ``Weak
bars'' lie mainly in ``Sc-Sd''. That is to say, both B/P and bar tend to
be strong with the earlier type galaxies, within the range from mid type
(Sb-Sbc) to late type (Sc-Sd). These results also support the idea that
B/P is bar seen side-on. 

These trends imply the evolution of disk galaxy from later type to
earlier type (e.g., \citet{b19}, \citet{b25}, \citet{b32}). 
That is to say, weak features including B/P or bar may
have formed firstly in Sc-Sd galaxy and then have grown up via secular
evolution of bulge, bar and spiral or some accretion events of
materials. Then the structures may become strong through the evolution
from Sc-Sd to Sb. If violent events like major merger happen, these
features may disappear and the galaxy may become an earlier type (S0-Sa
or elliptical) galaxy. The evolution would be affected with the
environment and thus the various type galaxies would exist in the
present universe. 

\subsection{Comparison with Previous Studies}

Previous observational studies in 1980s to 2000s have reported various
values of the B/P fraction; 1.2 percent \citep{b17}, over 13 percent
\citep{b10}, 20 percent \citep{b34}, and 45 percent {\citep{b12,b21}. 
The variety is caused by differences in sample selection, sample size,
analysis, and criteria to identify B/P structures, as \citet{b21} have pointed out. 

Our resulting B/P fractions against nearby edge-on galaxies are
about 18, 21 and 22 percent for each g, r and i-band, respectively. 
They are roughly intermediate between the previous papers. 
Our result is consistent with \citet{b34}, of which B/P fraction is 20
($\pm4$) percent. 
However, our result is seemingly inconsistent with the previous report
of significantly low fraction of 1.2 percent in \citet{b17} and
significantly high fraction of $45\pm8$ and $45\pm4.5$ percent in
\citet{b12} and \citet{b21}, respectively. 
The low fraction of 1.2 percent reported by \citet{b17} is the value
against the ``all'' disk galaxies of various inclination. Therefore the
fraction should rise when only edge-on galaxies are selected. 
On the other hand, the high fraction of B/P reported by \citet{b12} and
\citet{b21} would be due to including a very weak B/P feature. 

Here we compare directly bx+, bx and bx- in this study with type 1, type
2 and type 3 in \citet{b21}, respectively. 
The fractions of bx+, bx and bx- in this study are 1.5, 8.8 and 12.5
percent (i-band), respectively, while those of type 1, type 2 and type 3
in \citet{b21} are 4.1, 15.7 and 25.2 percent (using images of
B, V and R-band), respectively. 
Therefore the fractions for each class of our result are roughly 1/2 of
those of corresponding three classes of \citet{b21}. 
However, the criterion of B/P is somewhat strict in this study compared
to that of \citet{b21}; the bx- has not X-shape but boxy shape on the
residual images, while the type 3 is defined as ``close to box-shaped,
not elliptical'' on contour images of galaxies. 
It is possible that bx- should be corresponded not with type 3 but with
type 2, and that bx would be corresponded partly with type 1 and partly
with type 2. 
In fact, the total percentage of type 1 and type 2 is 19.8 percent,
which is very close to our result of total 22 percent for bx+, bx and
bx-. 
In addition, \citet{b34} has pointed out that the B/P fraction is about
20 percent neglecting very weak structure. 
Consequently we conclude that the B/P fraction in this study is consistent
with that of \citet{b34} and that of type 1 plus type 2 of \citet{b21}.

Our resulting bar fractions are about 54, 47 and 50 percent for each g,
r and i-band, respectively. 
Previously many authors have reported that the fraction of barred
galaxies is about 40 to 50 percent among all spiral galaxies in optical
bands. 
For example, the fraction of barred galaxies (classified as ``SB'' and
``SAB'') in RC3 \citep{b14} is about 43 percent over all spiral galaxies
(classified as ``S'').  
Recently \citet{b24} have reported that the fraction at $z\sim0$ is about
$44\pm7$ percent in optical and $60\pm7$ percent in NIR (near-infrared)
band based on 180 sample spirals observation. 
Moreover \citet{b3} have reported that it is about 48-52 percent at
$0.01\leq z \leq0.03$ from 3600 SDSS disk-dominated galaxies. 
Many authors (e.g., \citet{b15,b24}) have reported
that the bar fraction is considerably high in NIR band: about 60-70
percent. 
The reason of high bar fraction in NIR is generally recognized that the
faint bar obscured by dust is unveiled in NIR band observation.

Our result for bar fraction is nearly consistent with or slightly larger
than those of the above studies in optical bands. 
Since we use residual images to identify the bars, some faint bars are
easily found and thus it would give slightly larger fraction than those
of the previous studies.

As for the trend of B/P fraction versus Hubble type, \citet{b21} have
already found that the peak of B/P fraction (all of type 1, type 2 and
type 3) exists at Sb/Sbc and that of type 1 exist at Sa/Sab (see
Fig.2 in \citet{b21}). 
On the other hand, the peak of B/P fraction (all of bx+, bx and bx-) in
this paper exists at Sc in Fig.\ref{fig17} and that of strong B/P
(bx+) exist at Sb-Sbc in Fig.\ref{fig18}. 
The Hubble type transformed from the model $B/T$ in this paper tends to
be systematically later, as described in Section 4.1. 
However the tendency that the peak of stronger B/P shifts to earlier
type is commonly seen in Fig.18-21 of this paper and Fig.2 of
\citet{b21}. 
Hence we also conclude that the trend of B/P fraction against Hubble
type in this paper is generally consistent with that of \citet{b21}. 

Many authors (e.g., \citet{b20,b11,b35}) have presented the results of
two-dimensional bulge-disk decomposition. 
Since various models are used in these papers, we compare the results
with that of this study as a trend, instead of comparing directly the
obtained values of model parameters. 

For the model magnitude of total (bulge plus disk) or that of disk
component, they are roughly in good agreement with ``modelMag'' of SDSS,
which assumes only disk component. 
It indicates that the disk surface brightness is not sensitive to the
existence of bulge, other structures or the data quality, especially for
disk-dominated late type galaxies. 

On the other hand, bulge surface brightness in late type galaxy is
generally compact and is sensitive to the data quality. 
Sersic index of bulge, which is an important parameter to determin the
degree of concentration of bulge (effective radius and effective
luminosity), has been discussed by many authors. 
It is somewhat difficult to determin the value of index, because the
bulge profile is influenced by the data quality or the other components
like as dark lane. 
\citet{b20}, which have presented the result of decomposition for a
sample of 71,825 SDSS galaxies, have shown that the Sersic index ($1/n$)
of bulge is considerably influenced by the S/N or resolution (see Fig.6
of their paper). 
That is to say, the index $n$ tends to increase (i.e., $\beta$ decreases)
with degrading S/N, while it tends to decrease (thus $\beta$ increases)
with degrading resolution. 
\citet{b35} have also reported that the peak of distribution of
the index $n$ exists at about 0.5 to 1 (thus $\beta$ is 1 to 2) for 52,897
SDSS galaxies of which magnitude is around 16-18 mag in r-band (see
Fig.15 of their paper). 
Our result also shows that the $\beta=1$ (pure exponential bulge) is
generally well fitted, i.e., about 86-87 persent for edge-on data and
about 76-78 percent for face-on data have pure exponential bulge. 

In addition, \citet{b11} have reported that the bulge/disk
decomposition for 51 bright and large size galaxies, of which model and
fitting method are almost the same as this study. 
The average value of Sersic index $n$ in their paper is 3.52
($\beta=1/3.52$), while that of the above papers and this study is almost 1. 
However, the difference would be explained by the difference of Hubble
type and the data quality, i.e., most of the samples used in \citet{b11}
are early type galaxies and have relatively high resolution and large
size. 
Overall, our result of decomposition is thought to be approximately
consistent with these previous studies, as long as we use such SDSS data
with a selection bias. 

We also investigate various correlations between the values of surface
brightness parameters of bulge and disk, the colors, the absolute
magnitudes, the morphology and so on. The detail will be described in
the next paper II.

\section{Conclusions}

We have investigated statistically the nearby galaxies having the
Box/Peanut (B/P) and bar features, using the observational data
of edge-on and face-on nearby galaxies taken from SDSS (Sloan Digital
Sky Survey) DR7 archive. 1716 edge-on galaxies and 2689 face-on galaxies
are selected by the following condition: brighter than 17 magnitude in
r-band, larger than 5 arcsec of petrosian radius, and with axial ratios
smaller than 0.25 for edge-on or larger than 0.8 for face-on. They are
fitted with the model of 2-dimensional surface brightness of S\'{e}rsic
bulge and exponential disk. The following data have been finally
selected: 1253, 1312 and 1329 samples in g, r, and i-band in edge-on,
and 2042, 2020 and 1890 samples in g, r, and i-band in face-on,
respectively. The residual (observed minus model) images have been
produced to extract structures other than bulge and disk. The B/P, bar,
oval, ring, normal, elliptical, irregular, and merger galaxies in the
images are estimated by eye.

The main results are following: 

1. The catalogues containing surface brightness parameters and the
   morphology for both edge-on and face-on galaxies have been
   presented. 

2. We have found 292 B/P structures in the 1329 edge-on galaxies and 630
   bar structures in 1890 face-on galaxies in i-band. 

3. The B/P fraction is about 22 percent in the edge-on sample galaxies
   in i-band, and that of bar is about 33 percent in the face-on sample
   galaxies. After removing 639 elliptical galaxies from the face-on
   data, that of bar is about 50 percent. 

4. The fractions decrease with increasing the strength both for B/P and
   bar. 

5. The B/P fraction is generally roughly a half of those of bars,
   irrespective to the strength and the $B/T$. 

6. The strengths of B/Ps and bars are slightly correlated with $B/T$,
   i.e., the strong structure are found generally in the mid type
   (Sb-Sbc) galaxies rather than late type (Sc-Sd) galaxies.

We have discussed the difference between the fraction of B/Ps and that
of bars. The reason would be due to the viewing angle to the bar and/or
the inclination of galaxy. 
Considering these factors, our result supports the idea that the B/P is
bar seen side-on. Moreover we have compared our data with the previous
observational studies. The B/P fraction is roughly the medium of those
of previous studies, while the bar fraction is almost the same as many
of previous studies. 
Various values of B/P fractions among several studies are thought to
be mainly due to the criterion B/P feature.

\section*{Acknowledgments}

We thank Dr. R. L\"{u}tticke (referee of this paper) for careful reading
and constructive comments. 
Moreover we thank Dr. Yagi Yoshifumi for some helpful suggestions.

\section*{Appendix}

Here we show the figure of input-output scattering using model galaxy images
mentioned in Section 3.7. 

Fig.\ref{fig21}(a)-(i) indicate the possibility of fitting errors for
each parameter, assuming a condition similar to the observation limit
and the sample selection for the real SDSS galaxies. 

\begin{figure*}
\begin{center}
 
\includegraphics[width=170mm]{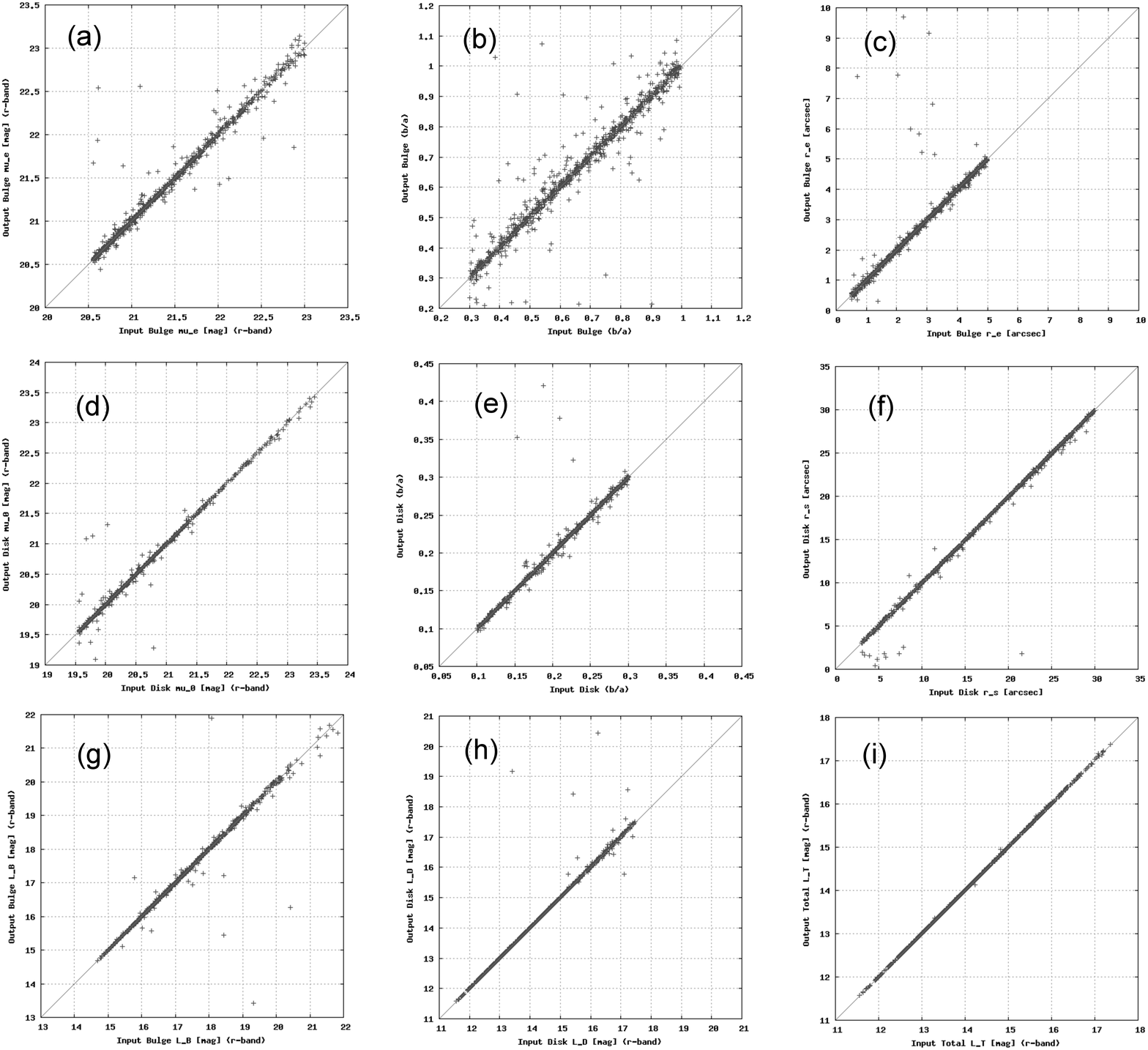}

\caption{The comparisons of input and outout values of parameters of models. Each panel show the
 reproducibility for the parameters of the randomly generated models in
 r-band as described in the text. The x-directions
 of each panel show the input values. The y-directions show those of
 output. (a)Bulge effective intensity $(\mu_e)_B
 \rm{[mag/arcsec^2]}$. (b)Bulge axial ratio $(b/a)_B$. (c)Bulge
 effective radius $(r_e)_B \rm{[arcsec]}$. (d)Disk central intensity
 $(\mu_0)_D\rm{[mag/arcsec^2]}$. (e)Disk axial ratio $(b/a)_D$. (f)Disk
 scale length $(r_s)_D\rm{[arcsec]}$. (g)Bulge integrated flux
 $m_B \rm{[mag/arcsec^2]}$. (h)Disk integrated flux $m_D
 \rm{[mag/arcsec^2]}$. (i)Total integrated flux $L_T
 \rm{[mag/arcsec^2]}$. As described in the text, the data having
 $(\mu_e)_B>23\rm{[mag/arcsec^2]}$ and $L_D>17.5\rm{[mag]}$ are excluded.}

\label{fig21}

\end{center}
\end{figure*}

\bsp

\label{lastpage}

\begin{landscape}
\begin{table}
 \begin{center}
\fontsize{6pt}{8.5pt}\selectfont
  \caption{Catalogue of edge-on galaxies in i-band. Only 10 data are
  shown in this table. Column (1): This catalogue number ``SDSS-i-eon-XXXX'' showing SDSS data, i-band,
  edge-on and serial number. Column (2)-(5): Data taken from SDSS
  DR7. Column (2): SDSS objID. Column (3) and (4): Right Ascention and
  Declination in degree unit (J2000.0). Column (5): redshift. Column (6)-(12):
  Parameter values obtained from model fitting. Column (6): Bulge
  effective intensity (flux) in $\rm mag/arcsec^2$ unit. Column (7): Bulge axial
  ratio. Column (8): Bulge effective radius in arcsec unit. Column (9):
  S\'{e}rsic law index $\beta$. Column (10): Disk central intensity in
  $\rm mag/arcsec^2$ unit. Column (11): Disk axial ratio. Column (12): Disk
  scale length in arcsec unit. Column (13): Bulge integrated flux in
  magnitude unit. Column (14): Disk integrated flux in magnitude
  unit. Column (15): Total integrated flux in magnitude unit. Column
  16: Bulge-to-total integrated flux ratio. Column (17): Reduced $\chi^2$ value for
  the model fitting to the object. Column (18): Morphology estimated by
  eye. bx+:Strong Box/Peanut. bx:Standard Box/Peanut. bx-:Weak
  Box/Peanut. br+:Strong bar. br:Standard bar. br-:Weak bar. nm:Normal
  disk galaxy. el:Elliptical galaxy. ir:Irregular
  galaxy. ov:Oval. rg:Ring. mg:Merger. Column (19): Hubble type
  transformed from the $B/T$. Sd: $B/T\leq0.05$. Sc: $0.05<B/T\leq0.1$. Sbc: $0.1<B/T\leq0.2$. 
  Sb: $0.2<B/Tleq0.3$. Sab: $0.3<B/T\leq0.4$. Sa/E (Sa or earlier): $0.4<B/T$. 
  Column (20): ``PGC'' galaxies catalogue number, if exist within 5 arcsec. Column (21): Distance from
  the object of SDSS to that of PGC in arcsec unit. 
}

   \label{tab1}

\begin{tabular}{@{}cccccccccccccccccrc@{}}
\hline
 number & objid & redshift & $(\mu_e)_B$ & $(b/a)_B$ & $(r_e)_B$ & S\'{e}rsic $\beta$ &
   $(\mu_0)_D$ & $(b/a)_D$ & $(r_s)_D$ & $m_B$ & $m_D$ & $m_T$ & $B/T$ & $\chi^2$ & morph. & H.type & PGC & dist. \\
(1) & (2) & (3) & (4) & (5) & (6) & (7) & (8) & (9) & (10) & (11) & (12)
& (13) & (14) & (15) & (16) & (17) & (18) & (19) \\
\hline 
SDSS-i-eon-0001 & 587722981744771128 & 0.010819 & 21.16 & 0.65 & 1.43 & 1.00 & 20.06 & 0.27 & 22.11 & 18.15 & 12.87 & 12.86 & 7.65e-03 & 1.18 & nm & Sd & 42747 & 0.6\\
SDSS-i-eon-0002 & 587722981745295552 & 0.02338 & 21.58 & 0.47 & 2.32 & 1.00 & 19.93 & 0.14 & 12.28 & 17.88 & 14.66 & 14.60 & 4.90e-02 & 1.12 & nm & Sd & 43198 & 1.7\\
SDSS-i-eon-0003 & 587722982292652134 & 0.052575 & 20.14 & 0.76 & 1.81 & 1.00 & 19.90 & 0.20 & 6.65 & 16.45 & 15.56 & 15.17 & 3.05e-01 & 2.62 & nm & Sab & 1137726 & 1.6\\
SDSS-i-eon-0004 & 587722982293176496 & 0.055175 & 21.54 & 0.95 & 2.74 & 0.75 & 19.93 & 0.13 & 9.08 & 16.62 & 15.42 & 15.11 & 2.50e-01 & 2.07 & nm & Sb & 91360 & 0.6\\
SDSS-i-eon-0005 & 587722982296912216 & 0.088005 & 21.12 & 1.01 & 1.06 & 1.00 & 20.27 & 0.18 & 5.27 & 18.28 & 16.61 & 16.40 & 1.76e-01 & 1.51 & nm & Sbc & 1134074 & 1.1\\
SDSS-i-eon-0006 & 587722982817333398 & 0.110171 & 21.82 & 0.73 & 1.69 & 1.00 & 20.50 & 0.13 & 5.45 & 18.35 & 17.13 & 16.83 & 2.47e-01 & 1.12 & nm & Sb & 1147239 & 1.5\\
SDSS-i-eon-0007 & 587722983355514897 & 0.072341 & 21.42 & 0.84 & 1.97 & 1.00 & 19.56 & 0.20 & 5.63 & 17.45 & 15.60 & 15.42 & 1.55e-01 & 1.77 & nm & Sbc & 1158721 & 4.4\\
SDSS-i-eon-0008 & 587722983356629128 & 0.022611 & 23.58 & 0.28 & 12.41 & 1.00 & 20.25 & 0.10 & 9.11 & 17.23 & 15.99 & 15.69 & 2.41e-01 & 1.72 & nm & Sb & - & -\\
SDSS-i-eon-0009 & 587722983368556772 & 0.02845 & 23.41 & 0.77 & 3.13 & 1.00 & 19.96 & 0.20 & 6.94 & 18.77 & 15.59 & 15.53 & 5.10e-02 & 2.44 & nm & Sc & 1159724 & 0.4\\
SDSS-i-eon-0010 & 587722983376093603 & 0.055354 & 22.15 & 1.05 & 1.64 & 1.00 & 21.04 & 0.13 & 8.08 & 18.37 & 16.84 & 16.60 & 1.96e-01 & 0.65 & bx- & Sbc & 92993 & 0.6\\
\hline
\end{tabular}
 \end{center}

 \begin{center}
\fontsize{6pt}{8.5pt}\selectfont
  \caption{Catalogue of face-on galaxies in i-band. Only 10 data are
  shown in this table. Column (1): This catalogue number
  ``SDSS-i-fon-XXXX'' showing SDSS data, i-band, face-on, and serial
  number. The other columns are the same as those of Table \ref{tab1}.}

   \label{tab2}

\begin{tabular}{@{}cccccccccccccccccrc@{}}
\hline
 number & objid & redshift & $(\mu_e)_B$ & $(b/a)_B$ & $(r_e)_B$ & S\'{e}rsic $\beta$ &
   $(\mu_0)_D$ & $(b/a)_D$ & $(r_s)_D$ & $m_B$ & $m_D$ & $m_T$ & $B/T$ & $\chi^2$ & morph. & H.type & PGC & dist. \\
(1) & (2) & (3) & (4) & (5) & (6) & (7) & (8) & (9) & (10) & (11) & (12)
& (13) & (14) & (15) & (16) & (17) & (18) & (19) \\
\hline 
SDSS-i-fon-0001 & 587722982280724639 & 0.077798 & 19.86 & 0.81 & 1.30 & 1.00 & 20.80 & 0.69 & 5.18 & 16.83 & 15.76 & 15.42 & 2.72e-01 & 1.78 & br-rg & Sb & 1135280 & 0.9\\
SDSS-i-fon-0002 & 587722982285312179 & 0.093877 & 20.57 & 0.83 & 1.02 & 1.00 & 20.83 & 0.74 & 4.89 & 18.03 & 15.86 & 15.72 & 1.19e-01 & 3.34 & brrg & Sbc & 1137394 & 2.6\\
SDSS-i-fon-0003 & 587722982292521047 & 0.054942 & 21.03 & 0.73 & 3.29 & 1.00 & 21.97 & 0.87 & 8.32 & 16.11 & 16.02 & 15.31 & 4.80e-01 & 0.60 & ovrg & Sa/E & 1137598 & 1.3\\
SDSS-i-fon-0004 & 587722982829981942 & 0.053721 & 22.81 & 0.31 & 3.11 & 1.00 & 20.03 & 1.01 & 3.94 & 18.97 & 15.15 & 15.12 & 2.88e-02 & 1.12 & nm & Sd & 51632 & 0.6\\
SDSS-i-fon-0005 & 587722983355383869 & 0.048184 & 19.42 & 0.87 & 2.05 & 0.75 & 21.27 & 1.01 & 10.32 & 15.17 & 14.50 & 14.03 & 3.51e-01 & 2.40 & ov & Sab & 1155146 & 0.4\\
SDSS-i-fon-0006 & 587722983363248237 & 0.029861 & 20.96 & 0.55 & 3.54 & 0.75 & 20.99 & 0.94 & 8.67 & 16.04 & 14.63 & 14.37 & 2.14e-01 & 5.75 & brmg & Sb & 49446 & 1.7\\
SDSS-i-fon-0007 & 587722983892648021 & 0.084617 & 21.19 & 0.47 & 3.10 & 0.75 & 21.03 & 0.82 & 5.05 & 16.73 & 15.92 & 15.50 & 3.21e-01 & 1.49 & br & Sab & 1169724 & 0.1\\
SDSS-i-fon-0008 & 587722983895990379 & 0.024142 & 21.83 & 0.89 & 1.29 & 1.00 & 20.55 & 0.84 & 6.79 & 18.71 & 14.74 & 14.71 & 2.51e-02 & 7.78 & br- & Sd & 46266 & 0.3\\
SDSS-i-fon-0009 & 587722983905100085 & 0.037183 & 21.17 & 0.89 & 1.13 & 1.00 & 20.27 & 0.90 & 6.00 & 18.34 & 14.61 & 14.57 & 3.12e-02 & 2.59 & br- & Sd & 52431 & 0.4\\
SDSS-i-fon-0010 & 587722984432402632 & 0.048132 & 22.10 & 0.42 & 3.05 & 0.75 & 19.55 & 0.85 & 3.48 & 17.81 & 15.08 & 14.99 & 7.47e-02 & 1.39 & br-rg & Sc & 3082174 & 0.7\\
\hline
\end{tabular}
 \end{center}

\end{table}
\end{landscape}

\begin{landscape}
\begin{table}
 \begin{center}
\fontsize{6pt}{8.5pt}\selectfont
  \caption{Catalogue of Box/Peanut shape galaxies in i-band extracted
  from the Table \ref{tab1}. Only 10 data are shown in this table. The columns
  are the same as those of Table \ref{tab1}. }
   \label{tab3}

\begin{tabular}{@{}cccccccccccccccccrc@{}}
\hline
 number & objid & redshift & $(\mu_e)_B$ & $(b/a)_B$ & $(r_e)_B$ & S\'{e}rsic $\beta$ &
   $(\mu_0)_D$ & $(b/a)_D$ & $(r_s)_D$ & $m_B$ & $m_D$ & $m_T$ & $B/T$ & $\chi^2$ & morph. & H.type & PGC & dist.\\
(1) & (2) & (3) & (4) & (5) & (6) & (7) & (8) & (9) & (10) & (11) & (12)
& (13) & (14) & (15) & (16) & (17) & (18) & (19) \\
\hline 
SDSS-i-eon-0010 & 587722983376093603 & 0.055354 & 22.15 & 1.05 & 1.64 & 1.00 & 21.04 & 0.13 & 8.08 & 18.37 & 16.84 & 16.60 & 1.96e-01 & 0.65 & bx- & Sbc & 92993 & 0.6\\
SDSS-i-eon-0014 & 587724198283575324 & 0.021312 & 19.91 & 1.07 & 1.23 & 1.00 & 19.09 & 0.22 & 7.25 & 16.69 & 14.44 & 14.31 & 1.12e-01 & 3.28 & bx+ & Sbc & 7114 & 1.4\\
SDSS-i-eon-0022 & 587724234257924238 & 0.04169 & 21.26 & 1.08 & 1.41 & 1.00 & 19.29 & 0.22 & 6.35 & 17.74 & 14.94 & 14.86 & 7.09e-02 & 0.87 & bx & Sc & 8373 & 0.8\\
SDSS-i-eon-0023 & 587724240153149551 & 0.03086 & 20.57 & 1.20 & 1.63 & 1.00 & 19.38 & 0.24 & 6.56 & 16.62 & 14.90 & 14.70 & 1.70e-01 & 1.51 & bx- & Sbc & 10704 & 1.1\\
SDSS-i-eon-0027 & 587724241232461942 & 0.065683 & 21.37 & 0.99 & 1.27 & 1.00 & 19.78 & 0.14 & 6.62 & 18.18 & 15.88 & 15.76 & 1.08e-01 & 1.39 & bx- & Sbc & - & -\\
SDSS-i-eon-0033 & 587725041700700283 & 0.077376 & 21.80 & 0.87 & 2.60 & 0.75 & 20.09 & 0.15 & 7.32 & 17.10 & 15.84 & 15.54 & 2.40e-01 & 0.92 & bx+ & Sb & 91140 & 1.7\\
SDSS-i-eon-0035 & 587725469056762056 & 0.039954 & 21.53 & 0.70 & 4.43 & 0.75 & 20.03 & 0.11 & 15.57 & 15.90 & 14.56 & 14.28 & 2.26e-01 & 1.80 & bx & Sb & 23609 & 1.9\\
SDSS-i-eon-0037 & 587725469599531113 & 0.042004 & 21.58 & 0.92 & 2.23 & 1.00 & 19.88 & 0.12 & 8.80 & 17.26 & 15.58 & 15.37 & 1.75e-01 & 3.11 & bx & Sbc & 2544317 & 1.7\\
SDSS-i-eon-0043 & 587725505558806667 & 0.028384 & 21.52 & 0.73 & 4.74 & 0.75 & 19.82 & 0.11 & 11.87 & 15.70 & 14.93 & 14.50 & 3.29e-01 & 3.91 & bx- & Sab & 91593 & 3.0\\
SDSS-i-eon-0045 & 587725551192113224 & 0.011182 & 18.72 & 0.60 & 3.47 & 0.75 & 19.36 & 0.20 & 23.22 & 13.74 & 12.34 & 12.07 & 2.16e-01 & 6.09 & bx & Sb & - & -\\

\hline
\end{tabular}
 \end{center}

 \begin{center}

\fontsize{6pt}{8.5pt}\selectfont

  \caption{Catalogue of barred galaxies in i-band extracted
  from the Table \ref{tab2}. Only 10 data are shown in this table. The columns
  are the same as those of Table \ref{tab2}. }
   \label{tab4}

\begin{tabular}{@{}cccccccccccccccccrc@{}}
\hline
 number & objid & redshift & $(\mu_e)_B$ & $(b/a)_B$ & $(r_e)_B$ & S\'{e}rsic $\beta$ &
   $(\mu_0)_D$ & $(b/a)_D$ & $(r_s)_D$ & $m_B$ & $m_D$ & $m_T$ & $B/T$ & $\chi^2$ & morph. & H.type & PGC & dist.\\
(1) & (2) & (3) & (4) & (5) & (6) & (7) & (8) & (9) & (10) & (11) & (12)
& (13) & (14) & (15) & (16) & (17) & (18) & (19) \\
\hline 
SDSS-i-fon-0001 & 587722982280724639 & 0.077798 & 19.86 & 0.81 & 1.30 & 1.00 & 20.80 & 0.69 & 5.18 & 16.83 & 15.76 & 15.42 & 2.72e-01 & 1.78 & br-rg & Sb & 1135280 & 0.9\\
SDSS-i-fon-0002 & 587722982285312179 & 0.093877 & 20.57 & 0.83 & 1.02 & 1.00 & 20.83 & 0.74 & 4.89 & 18.03 & 15.86 & 15.72 & 1.19e-01 & 3.34 & brrg & Sbc & 1137394 & 2.6\\
SDSS-i-fon-0006 & 587722983363248237 & 0.029861 & 20.96 & 0.55 & 3.54 & 0.75 & 20.99 & 0.94 & 8.67 & 16.04 & 14.63 & 14.37 & 2.14e-01 & 5.75 & brmg & Sb & 49446 & 1.7\\
SDSS-i-fon-0007 & 587722983892648021 & 0.084617 & 21.19 & 0.47 & 3.10 & 0.75 & 21.03 & 0.82 & 5.05 & 16.73 & 15.92 & 15.50 & 3.21e-01 & 1.49 & br & Sab & 1169724 & 0.1\\
SDSS-i-fon-0008 & 587722983895990379 & 0.024142 & 21.83 & 0.89 & 1.29 & 1.00 & 20.55 & 0.84 & 6.79 & 18.71 & 14.74 & 14.71 & 2.51e-02 & 7.78 & br- & Sd & 46266 & 0.3\\
SDSS-i-fon-0009 & 587722983905100085 & 0.037183 & 21.17 & 0.89 & 1.13 & 1.00 & 20.27 & 0.90 & 6.00 & 18.34 & 14.61 & 14.57 & 3.12e-02 & 2.59 & br- & Sd & 52431 & 0.4\\
SDSS-i-fon-0010 & 587722984432402632 & 0.048132 & 22.10 & 0.42 & 3.05 & 0.75 & 19.55 & 0.85 & 3.48 & 17.81 & 15.08 & 14.99 & 7.47e-02 & 1.39 & br-rg & Sc & 3082174 & 0.7\\
SDSS-i-fon-0011 & 587722984436990089 & 0.07035 & 20.61 & 0.86 & 0.88 & 1.00 & 20.38 & 0.81 & 3.82 & 18.34 & 15.76 & 15.66 & 8.47e-02 & 1.21 & brrg & Sc & 1178163 & 0.5\\
SDSS-i-fon-0013 & 587724198811271209 & 0.139565 & 20.91 & 0.67 & 1.51 & 1.00 & 21.80 & 0.77 & 9.23 & 17.76 & 16.02 & 15.82 & 1.67e-01 & 2.33 & br- & Sbc & - & -\\
SDSS-i-fon-0014 & 587724198822215805 & 0.148972 & 21.41 & 0.82 & 1.09 & 1.00 & 21.71 & 0.86 & 4.47 & 18.73 & 16.92 & 16.73 & 1.58e-01 & 0.92 & br- & Sbc & - & -\\
\hline
\end{tabular}
 \end{center}

\end{table}
\end{landscape}

\end{document}